\documentclass[a4paper,11pt]{article}
\usepackage{jheppub} 
\usepackage{lineno}
\usepackage{float}

\usepackage{comment}
\usepackage[normalem]{ulem}
\usepackage[english]{babel}
\usepackage{graphicx}
\usepackage{dcolumn}
\usepackage{bm}
\usepackage{blindtext}
\usepackage{verbatim}
\usepackage{mathrsfs}
\usepackage{musicography}
\usepackage{amsmath}
\usepackage{dsfont}
\usepackage{mathrsfs}
\usepackage{amssymb}
\usepackage{amsthm}
\usepackage{cancel}
\usepackage{physics}
\usepackage{epstopdf}
\usepackage{mathtools}
\usepackage{color}
\usepackage[usenames,dvipsnames]{pstricks}
\usepackage{epsfig}
\usepackage{pst-grad} 
\usepackage{pst-plot} 
\usepackage{hyperref}
\usepackage{verbatim}
\usepackage{afterpage}
\usepackage{sidecap}
\hypersetup{
    colorlinks=true,
    linkcolor=blue,
    filecolor=red,      
    urlcolor=cyan,
}
\usepackage{tikzsymbols}

\title{Gravitationally induced entanglement at finite temperature: A memory-driven time-crystalline phase?}

\author[1]{Mainak Dutta,}
\author[2,3]{Partha Nandi ,}
\author[1]{ Bibhas Ranjan Majhi}
\affiliation[1]{Department of Physics,\\ Indian Institute of Technology Guwahati, Guwahati 781039, Assam, India}
\affiliation[2]{Department of Physics, \\University 
	of Stellenbosch, Stellenbosch-7600, South Africa}
\affiliation[3]{National Institute for Theoretical and Computational Sciences (NITheCS),\\Stellenbosch, 7604, South Africa}

\emailAdd{d.mainak@iitg.ac.in} \emailAdd{pnandi@sun.ac.za} \emailAdd{bibhas.majhi@iitg.ac.in}

\abstract{We study the impact of thermal effects on gravity-induced entanglement (GIE) in a system of quantum harmonic oscillators interacting with classical linearly polarized gravitational waves (GWs). Specifically, we model the endpoints of interferometer arms in LIGO-like detectors as two-dimensional oscillators. Following the thermofield dynamics (TFD) approach, our analysis reveals that while thermal effects alone do not generate entanglement between independent oscillator modes, they serve as a catalyst, modifying the dynamical imprint of GWs. Notably, we identify a mixing of Bose-Einstein and Maxwell-Boltzmann distributions driven by thermal influences, which affects the statistical behavior of the quantum subsystem. Furthermore, gravitational interactions induce a quantum memory effect, leading to emergent periodic behavior in the reduced subsystem. This suggests a novel gravitationally induced breaking of time-translation symmetry, reminiscent of a prethermal time crystal (PTC). Our findings indicate that such effects could provide new theoretical insights into classical gravitational wave interactions.}

\begin{document}
\maketitle
\flushbottom

\section{Introduction}

The detection of gravitational waves by the LIGO and Virgo collaborations has revolutionized fundamental physics, providing an unprecedented opportunity to probe strong-field gravity and test general relativity in extreme astrophysical environments \cite{Abbott:2016blz}. Till then the
gravitational wave detection has become a major research focus, giving birth of multi-messenger astronomy. These waves originate from various astrophysical sources, including black holes \cite{Dokuchaev:2009}, neutron stars \cite{Lasky2015}, and even the early universe during the inflationary epoch \cite{Boyle:2008rg}. Such detections provide a unique window into fundamental physics, especially at timescales as short as $10^{-22}$ seconds after the Big Bang, where existing theories of gravity are expected to break down \cite{Kiefer2004}. The LIGO-Virgo-KAGRA (LVK) collaboration \cite{Aasi2015, Acernese2014, Aso2013, AsoKAGRA2013} has reported numerous gravitational wave events, primarily from binary black hole (BBH) mergers and binary neutron star (BNS) systems \cite{Abbott2023}. Future space-based observatories, such as DECIGO \cite{Seto2001, Kawamura2006, Yagi2011, Yagi2017} and LISA, will probe lower-frequency regimes, while atom interferometers and pulsar timing arrays (PTAs) \cite{QuantumSciTech2021, Kramer2013, Babak2016, Lentati2015, Shannon2015, Manchester2013, Hobbs2010, Verbiest2016, Hazboun2018} will extend sensitivity even further.

Beyond their astrophysical significance, gravitational waves offer a unique platform to explore the interface between quantum mechanics and gravity \cite{PhysRevLett.126.041302}. A central question in modern theoretical physics is whether gravity itself must be quantized or whether classical gravity can still induce quantum correlations \cite{PhysRevLett.47.979}. In this context, recent studies suggest that tree-level quantum effects may be detectable even within linearized gravity, offering valuable insights into this fundamental issue. An emerging research direction following the success of gravitational wave detection is the search for quantum signatures within these waves, particularly the elusive detection of gravitons \cite{PhysRevLett.127.081602, PhysRevD.109.096023}. While detecting individual gravitons remains a formidable challenge \cite{PhysRevD.104.046021}, some researchers question whether gravity must be fully quantized \cite{PhysRevLett.75.1260}. Despite these debates, ongoing studies seek experimental signatures that might provide indirect evidence, with some predictions suggesting that future detectors could achieve the required sensitivity \cite{Tobar2024}. 

However, unlike electromagnetism—where the quantization of matter necessarily implies the quantization of the electromagnetic field—the detection of gravitational wave noise or discrete detector responses does not, by itself, rule out a classical interpretation of gravity \cite{Clauser1974, PhysRevD.109.044009}. This ambiguity motivates alternative approaches to probing the quantum nature of gravity, such as studying gravity-induced entanglement.

One approach to addressing this question is through the study of \textit{gravity-induced entanglement} \cite{Bose2017, Marletto2017,Nandi:2024jyf,Nandi:2024zxp,Ruggiero:2024pzv,Kaku:2024lgs,Jones:2024npd}. If two massive objects, interacting solely via Newtonian gravity, evolve from an initially separable state into an entangled state, it would indicate that their gravitational interaction cannot be described as a local operation and classical communication (LOCC), suggesting that gravity must be inherently quantum \cite{Bose2017, Marletto2017}. However, some arguments challenge the idea that Newtonian entanglement directly implies quantization of the gravitational field \cite{Martin-Martinez:2023}. Indeed, gravity-induced entanglement alone does not always  necessarily serve as conclusive proof of quantum gravitational effects. In our recent work \cite{Nandi:2024jyf}, we demonstrated that entanglement harvesting can serve as evidence for the quantum nature of gravity, specifically in scenarios where gravitational waves interact with individual quantum detector modes without making any classical coupling between them. This requires consideration of the quantum nature of gravitational waves. However, in cases where gravitational waves mediate a direct classical interaction between detector modes, entanglement can still arise purely from classical gravitational effects \cite{Nandi:2024zxp}.


While much of the discussion surrounding gravitational waves focuses on their potential connection to quantum gravity, there is also growing interest in understanding how classical gravitational waves influence quantum systems \cite{Speliotopoulos1995, PhysRevD.108.124069}. Instead of probing the quantum nature of gravity itself, experiments such as the COW experiment \cite{Colella1975} and studies by Pikovski et al. \cite{Pikovski2015} investigate how quantum particles behave in external classical gravitational fields. Phenomena associated with quantum field theory in curved spacetime—such as Hawking radiation—are typically considered relevant only in extreme conditions, such as high-energy regimes and strong gravitational fields. However, it remains an open question whether quantum mechanics can introduce novel effects even at lower energies when interacting with gravitational waves.

In realistic experimental settings, quantum systems are never perfectly isolated. Thermal noise is an unavoidable factor that can influence quantum coherence and entanglement dynamics. Given the potential for gravitational waves to affect these quantum properties, it is crucial to explore how such effects manifest in finite temperature environments. Gravitational wave detection in LIGO interferometers occurs at extraordinarily small length scales—on the order of $10^{-18}$ m—where quantum effects on the detector become significant \cite{Caves1980, Thorne1997}. Additionally, resonant bar detectors have achieved remarkable sensitivity \cite{Dimopoulos2008}, tracking test mass positions at precision levels of $10^{-20}$ m. Understanding quantum effects at these scales requires a careful examination of how classical gravitational waves interact with quantum mechanical systems under the influence of external thermal environments.\\

Building on these developments, we investigate how quantum effects induced by classical gravitational waves on a quantum mechanical system are influenced by an external thermal environment—an inevitable feature of real-world experiments. Specifically, we focus on the following key aspects:

\begin{enumerate}
    \item Exploring the effects of linearized gravity on a quantum system in the presence of environmental temperature.
    \item Distinguishing between the roles of environmental temperature and the background gravitational effects on quantum entanglement.
    \item Investigating the emergence of periodic entanglement dynamics and potential signatures of time crystal like  behavior in gravitationally coupled quantum systems.

\end{enumerate}

In our recent work \cite{Nandi:2024zxp}, we partially addressed the first aspect by analyzing a quantum system composed of two-dimensional harmonic oscillators, modeling the endpoints of a LIGO-like detector. We investigated their response to gravitational wave perturbations, particularly in the context of entanglement harvesting via classical gravitational waves, assuming an idealized scenario without environmental effects. 
We particularly concentrate on the linearly polarized GWs.

To incorporate thermal effects into the quantum harmonic oscillator system, we prepare an initial thermal vacuum state, described by a pure density matrix. Instead of using a conventional mixed-state formalism, we employ the thermofield dynamics (TFD) framework \cite{Das:1997gg,Takahashi:1996zn}, which introduces fictitious thermal modes alongside the real oscillator modes. This approach enables us to capture the interplay between thermal fluctuations and gravitational interactions effectively.

While thermal effects alone do not generate entanglement between independent oscillator modes, gravitational waves facilitate entanglement harvesting, with thermality modifying its characteristics. This modulation can be quantitatively analyzed through the purity function, obtained by tracing out unobserved modes. Notably, entanglement extends beyond the physical modes, encompassing the auxiliary thermal modes introduced in the TFD framework. However, this entanglement cannot be entirely erased by thermal effects, underscoring the robustness of gravitationally mediated quantum correlations.

The reduced subsystem of a thermally prepared quantum system exhibits a periodic time dependence distinct from the original system Hamiltonian. This periodicity arises from gravitational wave interactions, leading to the dynamical evolution of entanglement and a quantum memory effect that persists even in the absence of ongoing gravitational interactions. While classical gravitational wave memory effects are well studied, their quantum counterpart remains largely unexplored. Our findings suggest that quantum memory may preserve gravity-induced quantum information over time, despite the presence of thermal noise.

Additionally, the quantum memory effect manifests in the statistical properties of energy observables within the gravitationally perturbed reduced density matrix, incorporating both environmental thermal effects and gravitational influences. The reduced subsystem behaves as a canonical ensemble, exchanging energy with its surroundings. While thermalization due to the environment leads to a statistical mixture, gravitational wave interactions induce an additional form of thermalization, purely driven by gravitational effects. Their interplay further modifies the system’s statistical properties, resulting in stimulated thermalization.

Our analysis reveals that the quantum memory effect induced by gravitational waves leads to discrete time-translation symmetry breaking in the reduced subsystem, resulting in emergent periodic behavior. This behavior shares similarities with prethermal time crystals (PTCs) \cite{PhysRevLett.117.090402}, which resist thermalization over extended timescales. However, unlike conventional PTCs, which rely on slow heating from Floquet driving, the periodic response in our system is sustained solely by gravitational wave interactions. Notably, this analogy is valid only in the presence of a finite background temperature. While the quantum memory effect persists even in the absence of thermal effects, its interpretation as a prethermal time-crystal-like phase is meaningful only in a thermal environment, where the system exhibits an unconventional mixture of Bose-Einstein and Maxwell-Boltzmann distributions.\\

The paper is structured as follows: Section \ref{Sec2} introduces the quantum mechanical model, where 2D harmonic oscillators interact with linearly polarized classical gravitational waves in the transverse-traceless (TT) gauge. The effective Hamiltonian governing the system is derived, and the thermofield dynamics (TFD) approach is introduced to incorporate finite-temperature effects.  Section \ref{Sec3} presents our main results, including the analysis of the reduced density matrix, statistical properties, and entanglement measures such as purity and entropy.  Section \ref{Sec4} explores the emergence of a gravitationally induced time-modulated quantum state, highlighting the notion of a memory-retaining time-crystal-like phase.  
Section \ref{Sec5} concludes with a discussion of the broader implications of our findings and outlines potential future research directions. Finally, the appendices provide detailed derivations, including the computation of the reduced density matrix, explicit formulations of entanglement measures, and supporting technical analyses.

\section{The setup and Hamiltonian}\label{Sec2}

\subsection{System Hamiltonian}

In this section, we investigate how classical gravitational waves (GWs) influence a quantum detector mass, without addressing the quantum nature of GWs themselves. According to Einstein’s equivalence principle, a single free particle does not experience direct effects from GWs in a classical setting. However, GWs induce geodesic deviations, leading to relative accelerations between nearby massive objects \cite{Misner:1973prb}. This implies that detecting GW effects requires at least two masses and an analysis of their relative motion.

GWs primarily induce displacements perpendicular to their propagation. In the linearized regime, this can be modeled as the motion of two particles along transverse directions \cite{Maggiore:2007ulw}. To ensure a well-defined quantum mechanical treatment, we assume these particles are confined within independent harmonic potentials. This confinement prevents wavefunction spreading and mimics the restoring forces acting on suspended mirrors in interferometric detectors like LIGO. In these setups, GWs induce minuscule displacements in the mirrors’ center of mass ($\delta L \sim 10^{-18} \text{m}$), significantly smaller than the associated de Broglie wavelength ($\lambda_D$), assuming the mirrors behave effectively as point masses \cite{PhysRevD.103.044017}. This suggests that a purely classical treatment may be inadequate, highlighting the need to explore quantum effects in high-precision GW detection.


To describe the quantum mechanics of test masses obeying the geodesic deviation equation with an additional oscillatory restoring force, we adopt a Hamiltonian formulation. In the weak-field limit, assuming a GW propagating along the \( z \)-axis, the system is governed by
\begin{equation}
H(t) = \sum_{j=1,2} \left( \frac{p_j^2}{2m} + \sum_{k=1,2} \Gamma^j_{0k}(t) x^k p_j + \frac{1}{2} m\omega^2 x_j^2 \right),
\label{x1}
\end{equation}
where \( x_j \) denotes particle displacements, \( m \) represents the mirror mass, and \( \Gamma^j_{0k}(t) = \frac{1}{2} \dot{h}_{jk}(t) \) encodes the GW perturbation \cite{Maggiore:2007ulw}. The parameter \( \omega \) corresponds to the harmonic potential frequency, providing a controlled setting for examining quantum effects in GW detection, an approach aligned with previous studies \cite{Speliotopoulos1995,PhysRevD.108.124069}. The above Hamiltonian represents two independent one-dimensional Harmonic oscillators which are interacting with each other through the background GWs. A detailed derivation of this Hamiltonian from first principles is presented in Appendix \ref{Apprev1}.

To simplify the analysis, we introduce a suitable canonical transformation in phase space, which allows us to express the plus polarization in a form where it effectively behaves like the cross polarization, reducing the complexity of the interaction (see, \cite{Nandi:2024jyf} for details). After performing this canonical transformation, the Hamiltonian takes the following form:
\begin{equation}
H(t) = \sum_{j=1}^{2} \left(\alpha p_j'^2 + \beta x_j'^2 \right) + g(t) \left(x_1' p_2' + x_2' p_1'\right),
\label{H2}
\end{equation}
where \( \alpha = \frac{1}{2m} \), \( \beta = \frac{1}{2}m\omega^2 \), and \( g(t) = \pm\sqrt{\delta^2(t) + 4\gamma^2(t)} \) describes the interaction with the GWs. Here $\delta$ and $\gamma$ correspond to cross and plus polarizations of GW, respectively. For self-sufficiency of the above discussion, we present a detailed description in Appendix \ref{App1}.
It is advantageous to introduce annihilation (corresponding creation) operators (with $\hbar=1$)
\begin{equation}
\hat{a}_i = \left(\frac{\alpha}{\beta}\right)^{1/4} \left( \frac{\sqrt{\frac{\beta}{\alpha}} \hat{x}_i' + i \hat{p}_i'}{\sqrt{2}} \right),
\label{a_i}
\end{equation}
for the two-mode harmonic oscillator, satisfying the commutation relation $[\hat{a}_i, \hat{a}_j^\dagger] = \delta_{ij}$. These allow us to represent the system in the form of quadratic Hermitian expressions as follows:
\begin{equation}
\hat{H}(t) = 2 \sqrt{\alpha \beta} \left( \sum_i \hat{N}_i + 1 \right) + i g(t) (\hat{a}_1^\dagger \hat{a}_2^\dagger - \hat{a}_1 \hat{a}_2),
\label{H3}
\end{equation}
where \( \hat{N}_i = \hat{a}_i^\dagger \hat{a}_i \) is the number operator. The GW interaction introduces a two-mode squeezing term through \( g(t) \), and this squeezing effect significantly influences the quantum states of the system.

To further explore the underlying symmetry algebra of the system Hamiltonian in the presence of gravitational wave interactions and enhance the clarity of our computations, we introduce the following bilinear operators:
\begin{equation}
\hat{S}_0 = \frac{1}{2} \left( \sum_i \hat{N}_i + 1 \right), \quad \hat{S}_{-} = \hat{a}_1 \hat{a}_2, \quad \hat{S}_{+} = \hat{a}_1^\dagger \hat{a}_2^\dagger.
\label{S_0}
\end{equation}
These operators satisfy the commutation relations:
\begin{equation}
[\hat{S}_0, \hat{S}_\pm] = \pm \hat{S}_\pm, \quad [\hat{S}_+, \hat{S}_-] = -2 \hat{S}_0,
\label{SU_comm}
\end{equation}
demonstrating that they form a closed algebra under $SU(1,1)$. Thus, they serve as the generators of the $SU(1,1)$ group. Expressing the Hamiltonian in terms of these generators, we obtain:
\begin{equation}
\hat{H}(t) = \Omega \hat{S}_0 + ig(t) (\hat{S}_+ - \hat{S}_-),
\label{H4}
\end{equation}
where $\Omega = 4\sqrt{\alpha\beta} = 2\omega$. This formulation explicitly reveals the $SU(1,1)$ structure of the Hamiltonian, with the properties $\hat{S}_0^{\dagger} = \hat{S}_0$, $\hat{S}_+^{\dagger} = \hat{S}_-$, and $\hat{S}_-^{\dagger} = \hat{S}_+$. 

\subsection{Thermal vacuum states}
To encounter both thermal effects and the influence of gravitational waves (GWs), we consider a system of two-dimensional quantum harmonic oscillators (HOs) initially prepared in a thermal vacuum state, following the thermofield dynamics (TFD) approach. This formulation introduces an auxiliary ``tilde" system corresponding to each physical HO, allowing finite-temperature systems to be treated as zero-temperature systems within an extended Hilbert space. In this framework, the thermal density matrix is represented as a pure state in the enlarged Hilbert space \cite{Biswas1989, Das:1997gg,Takahashi:1996zn}.

To incorporate thermal effects, we introduce a fictitious tilde system. Within the thermofield dynamics framework, the total Hamiltonian is given by
\begin{equation}
\mathcal{H}(t) = \mathcal{H}_0 + \mathcal{H}_{\text{int}}(t),
\label{eq:thermal_hamiltonian}
\end{equation}
where the unperturbed extended  Hamiltonian is
\begin{equation} \mathcal{H}_0 = H_0 - \widetilde{H}_0 = \sum_{i=1}^{2} \omega \left( a_i^\dagger a_i - \tilde{a}_{i}^\dagger \tilde{a}_{i} \right),
\end{equation}
and the extended interaction Hamiltonian due to gravitational waves is
\begin{equation} \mathcal{H}_{\text{int}}(t) = H_{\text{int}}(t) - \widetilde{H}_{\text{int}}(t) = ig(t) \left( a_1^\dagger a_2^\dagger - a_1 a_2 \right) - ig(t) \left( \tilde{a}_1^\dagger \tilde{a}_2^\dagger - \tilde{a}_1 \tilde{a}_2 \right).
\label{int}
\end{equation} 
Here, \(\mathcal{H}_{\text{int}}\) represents the interaction due to gravitational waves. The thermal vacuum state associated with the unperturbed extended Hamiltonian \(\mathcal{H}_0\) can be written as (with $\hbar=1$):
\begin{equation} 
    \ket{0,0;\beta,t=0} = Z^{-1/2}(\beta)\sum^\infty_{n_1,n_2=0} e^{-\beta(n_1 + n_2 + 1)\omega/2} 
    \ket{n_1, n_2, \widetilde{n}_1, \widetilde{n}_2}. 
    \label{pure}
\end{equation}
Although the thermal vacuum state in Eq.~($\ref{pure}$) is globally pure, its reduced density matrix in the physical sector alone is mixed, reflecting the thermal nature of the initial state. This viewpoint is central to the TFD approach, which not only reproduces the thermal density matrix correctly but also allows one to analyze finite-temperature quantum dynamics using operator methods in a doubled Hilbert space.
More precisely, TFD provides a unified, operatorial, and fully unitary framework for treating quantum systems at finite temperature. Rather than working with a mixed-state density matrix, the thermal state is embedded as a pure state entangled between physical and fictitious (tilde) sectors:
$\ket{0,0;\beta,t=0} \in \mathbf{H} \otimes \tilde{\mathbf{H}}$,
and the physical thermal density matrix is obtained by tracing over the tilde degrees of freedom:
\begin{equation}
\rho_\beta = \mathrm{Tr}_{\tilde{\mathbf{H}}} \left[ \ket{0,0;\beta}\bra{0,0;\beta)} \right].
\end{equation}
Thus, although the full state evolves unitarily under the extended Hamiltonian \( \mathcal{H}(t) \), the reduced dynamics in the physical sector effectively capture the influence of thermal fluctuations.

This structure becomes especially useful in our setup, where two types of coupling are simultaneously present:
\begin{itemize}
    \item[(i)] The gravitational wave induces coherent coupling and entanglement between the two physical oscillators via a time-dependent two-mode squeezing interaction (\ref{int}).
    \item[(ii)] The thermal background is encoded as preexisting entanglement between each physical oscillator and its corresponding fictitious thermal partner.
\end{itemize}
The TFD formalism enables us to algebraically isolate and track both of these entanglement channels in a transparent and unified manner. As a result, we can identify how temperature modifies—but does not itself generate—the gravitationally induced entanglement between physical oscillators. In this sense, we will see that the temperature acts as a \emph{catalyst} for quantum correlations driven by spacetime geometry, a central theme in our analysis that would be difficult to extract from conventional thermal-state formalisms.

Here, we assume that both harmonic oscillators are at the same finite temperature, given by $k_{B}T=1/\beta$.
The partition function is given by:
\begin{equation}\label{partition_function}
    Z(\beta) = \sum_{n_1,n_2=0}^\infty e^{-\beta E_{n_1,n_2}} = \frac{e^{-\beta\omega}}{(1-e^{-\beta\omega})^2},
\end{equation}
where this state resides in the Hilbert space $\mathcal{H}= \mathbf{H} \otimes \widetilde{\mathbf{H}}$. The space \(\mathbf{H}\) is defined as:
\begin{equation}
\mathcal{H} = \text{span}\left\{|n_1, n_2; \widetilde{m}_1, \widetilde{m}_2\rangle = |n_1, n_2\rangle \otimes \widetilde{|m_1, m_2\rangle}\right\}~.
\label{eq:hilbert_space}
\end{equation}
In the above \(\ket{n_1, n_2, \widetilde{n}_1, \widetilde{n}_2}\) denotes the non-thermal eigenstates of the 2D harmonic oscillator at \(t = 0\), with energy \(E_{n_1,n_2}=(n_1+n_2+1)\omega\) and \(n_1,n_2=0,1,2,\cdots\). The parameter \(\omega\) represents the classical frequency of the 2D isotropic oscillator. In the next section, we study the effect of gravitational waves on this thermal ground state within the interaction picture.

\section{Entanglement Phenomenon}\label{Sec3}

\subsection{Interaction Picture Evolution}
We consider the interaction between the modes of the harmonic oscillator (HO) and gravitational waves (GWs) to be weak. This allows us to treat the second term in \eqref{eq:thermal_hamiltonian} as a perturbation. To examine the effect of gravitational interaction on the thermal state of the detector prepared at time \(t=0\), we express it as $\ket{0,0;\beta,t}_{I} = U_I(t,0) \ket{0,0;\beta,t=0}$, where 
\begin{equation}\label{eq:unitary_operator}
    U_I(t,0) = \text{T} \exp\left(-i \int_0^t \mathcal{H}^I_{\text{int}}(t') dt'\right),
\end{equation}
with \( \text{T} \) denotes the product ordered in time. In addition, we have 
\begin{equation}\label{eq:interaction_hamiltonian}
    \mathcal{H}^I_{\text{int}}(t) = e^{i\mathcal{H}_0 t} \mathcal{H}_{\text{int}}(t) e^{-i\mathcal{H}_0 t}.
\end{equation}
To proceed further using Baker-Campbell-Hausdorff  (BCH) lemma the above one is expressed in the following form:
\begin{equation}\label{3.3}
    \mathcal{H}^I_{\text{int}}(t) = ig(t) \Big\{ (e^{i\Omega t}S_+-e^{-i\Omega t}S_-) + (e^{i\Omega t}\tilde{S}_--e^{-i\Omega t}\tilde{S}_+) \Big\}.
\end{equation}
Further the time order product in (\ref{eq:unitary_operator}) can be expanded using the Magnus expansion formula. Keeping the terms up to the second order in perturbation, we obtain
\begin{eqnarray} \label{3.4}
    U_I(t,0) &=& 1 - i \int_0^t \mathcal{H}^I_{\text{int}}(t_1) dt_1 
     + \frac{(-i)^2}{2!} \int_0^t dt_1 \int_0^{t_1} dt_2 
    \left[\mathcal{H}^I_{\text{int}}(t_1), \mathcal{H}^I_{\text{int}}(t_2)\right] 
     \nonumber
     \\
     &+& \frac{(-i)^2}{2!} 
    \bigg( \int_0^t \mathcal{H}^I_{\text{int}}(t_1) dt_1 \bigg)^2 + \mathcal{O}(g^3).
\end{eqnarray}
Now it is important to note that the primitive commutator of the interaction Hamiltonian at two different times is expressed as follows:
\begin{equation}\label{3.5}
    \left[\mathcal{H}^I_{\text{int}}(t_1), \mathcal{H}^I_{\text{int}}(t_2)\right] = 
    -4ig(t_1)g(t_2)S_0\sin[{\Omega(t_1-t_2)}] 
    + 4ig(t_1)g(t_2)\tilde{S}_0\sin[{\Omega(t_1-t_2)}].
\end{equation}
Then by considering terms up to \(O(g^2)\), we derive the following expression:
\begin{eqnarray}
\label{A2}
    U_I(t,0) &=& 1 + i(c_0 S_0 - c_+ S_+ + c_- S_-) + i(c_- \tilde{S}_+ - c_+ \tilde{S}_- -c_0 \tilde{S}_0) 
    \nonumber
    \\
    &-& \frac{1}{2}(c_+ S_+ - c_- S_- + c_+ \tilde{S}_- - c_- \tilde{S}_+)^2~,
\end{eqnarray}
with 
\begin{align}
    c_+(t)&=i\int^t_0 dt_{1} g(t_1)e^{i\Omega t_1}~;\notag\\
    c_-(t)&=i\int^t_0 dt_{1} g(t_1)e^{-i\Omega t_1}=-\overline{c}_+(t)~; \notag\\
    c_0(t)&=2\int^t_0dt_1\int^{t_1}_0dt_2g(t_1)g(t_2)\sin[{\Omega(t_1-t_2)}]~.
    \label{j}
\end{align}
For completeness, steps to obtain (\ref{A2}) is presented in Appendix \ref{App2}.


Using the above results we find the evolution of the thermal ground state (in interaction picture) at later time \( t \) as
\begin{eqnarray}
\label{A3}
    \ket{0,0;\beta,t}_I &=& (1 - e^{-\beta\omega})\sum_{n_x,n_y=0}^\infty e^{-\frac{\beta \omega}{2}(n_1+n_{2})} \Bigg[ \ket{n_1,n_2,\tilde{n}_1,\tilde{n}_2} 
    - i(1 + e^{-\beta\omega})\sqrt{(n_1+1)(n_2+1)}
    \nonumber
    \\
    &\times&\Big( c_{+} \ket{n_1+1,n_2+1,\tilde{n}_1,\tilde{n}_2} - c_{-} \ket{n_1,n_2,\tilde{n}_1+1,\tilde{n}_2+1} \Big)
    \nonumber
    \\
    &-& \frac{(1 + e^{-\beta\omega})^2}{2} \sqrt{(n_1+1)(n_1+2)(n_2+1)(n_2+2)}
    \Big( c_+^2 \ket{n_1+2,n_2+2,\tilde{n}_1,\tilde{n}_2}  
    \nonumber
    \\
     &+&  c_-^2 \ket{n_1,n_2,\tilde{n}_1+2,\tilde{n}_2+2} \Big) 
    \nonumber
    \\
    &+& c_{+}c_{-}\Big\{ \Big(n_1 n_2 + (1 + e^{-\beta\omega})(n_1+1)(n_2+1)\Big)\ket{n_1,n_2,\tilde{n}_1,\tilde{n}_2} 
    \nonumber
    \\
    &+& (n_1+1)(n_2+1) \ket{n_1+1,n_2+1,\tilde{n}_1+1,\tilde{n}_2+1} \Big\} \Bigg].
\end{eqnarray}
Here, we have utilized the principles of second-order time-dependent perturbation theory to derive the above state. A detailed derivation is outlined in Appendix \ref{App2}. To obtain the reduced-density matrix of a thermal oscillator, specifically, the first oscillator, we trace over the degrees of freedom associated with the second oscillator and its corresponding tilde system. This procedure results in the following:
\begin{equation}\label{eq:reduced_density_matrix_initial}
\rho_{1,\tilde{1}}^\beta(t) = \mathrm{Tr}_{2,\tilde{2}}\left( \ket{0,0;\beta,t}{_I}~ {_I}\bra{0,0;\beta,t} \right)~.
\end{equation}
After keeping the terms up to the second order in the interaction \(g(t)\), the reduced density matrix is given by:
\begin{equation}\label{eq:reduced_density_matrix}
\begin{aligned}
    \rho_{1,\Tilde{1}}^\beta(t) \approx & \sum_{n_1,n'_1=0}^\infty e^{-\frac{\beta\omega}{2}(n_1+n'_1)} \Bigg[ 
    \Bigg\{1 - e^{-\beta\omega} - \abs{c_+}^2 \Big((1 + e^{-\beta\omega})
    (n'_1+1) + n'_1 e^{-\beta\omega}\\
     &+ (1 + e^{-\beta\omega})(n_1+1)+ n_1 e^{-\beta\omega}\Big)\Bigg\} 
    \ket{n_1,\Tilde{n}_1} \bra{n'_1,\Tilde{n}'_1}\\ 
      & + \abs{c_+}^2 (1 + e^{-\beta\omega})^2 \sqrt{(n_1+1)(n'_1+1)} 
     \Bigg( \ket{n_1,\Tilde{n}_1+1} \bra{n'_1,\Tilde{n}'_1+1}\\
    & 
    + \ket{n_1+1,\Tilde{n}_1} \bra{n'_1+1,\Tilde{n}'_1} \Bigg)  - \abs{c_+}^2 e^{-\beta\omega}(n_1 + n'_1 + 2) 
     \ket{n_1+1,\Tilde{n}_1+1} \bra{n'_1+1,\Tilde{n}'_1+1} 
    \Bigg]~.
\end{aligned}
\end{equation}
Since this state is constructed by tracing the information from the second oscillator and its fictitious counterpart, it represents the {\it evolved thermal vacuum state} of the first HO. Next, by tracing out the fictitious  mode of the first oscillator tilde system and focusing only on the  physical harmonic oscillator (HO), we obtain the reduced density matrix:
\begin{equation}
    \rho^{\beta}_{1}(t) = \text{Tr}_{\tilde{1}} \rho^{\beta}_{1,\tilde{1}} = \rho^{\beta}_{1} (0) + \delta \rho^{\beta}_{g}(t)~,
    \label{gh}
\end{equation}
where \(\rho^{\beta}_{1}(0) = \sum_{n_{1}=0}^\infty A_{n_{1}} \ket{n_{1}}\bra{n_{1}}\) represents the initial unperturbed thermal density matrix of the first HO. This unperturbed part is a mixed state while unperturbed part of (\ref{eq:reduced_density_matrix}) is a pure state.  Therefore we recognize (\ref{gh}) as the {\it evolved thermal state} of the first HO. In the above, the correction due to the gravitational wave interaction is given by:
\begin{equation}
    \delta \rho^{\beta}_{g}(t) = \sum_{n_{1}=0}^\infty \left( |c_+|^2 B_{n_{1}} \ket{n_{1}} \bra{n_{1}} + |c_+|^2 C_{n_{1}} \ket{n_{1}+1} \bra{n_{1}+1} \right)~,
    \label{d}
\end{equation}
where the coefficients \(A_{n_{1}}\), \(B_{n_{1}}\), and \(C_{n_{1}}\) are defined as follows,
\begin{eqnarray}
&&A_{n_{1}} = e^{-n_{1}\beta\omega}(1 - e^{-\beta\omega})~;
\label{eq:coefficient_A}
\\
&&B_{n_{1}} = e^{-n_{1}\beta\omega} \Big[-2(1 + e^{-\beta\omega})(2n_{1}+1) + (1 + e^{-\beta\omega})^2(n_{1}+1)\Big]~;
\label{eq:coefficient_B}
\\
&&C_{n_{1}} = e^{-n_{1}\beta\omega}(1 + e^{-\beta\omega})^2(n_{1}+1)~.
\label{eq:coefficient_C}
\end{eqnarray}
In the limit as \(n_{1}=0\) and \(\beta \to \infty\), equation \eqref{gh} reduces to the result previously obtained in \cite{Nandi:2024zxp}.

\subsection{Loss of Purity and its role in quantum entanglement}
\begin{figure}[htbp!]
\centering
\includegraphics[width=0.6\linewidth]{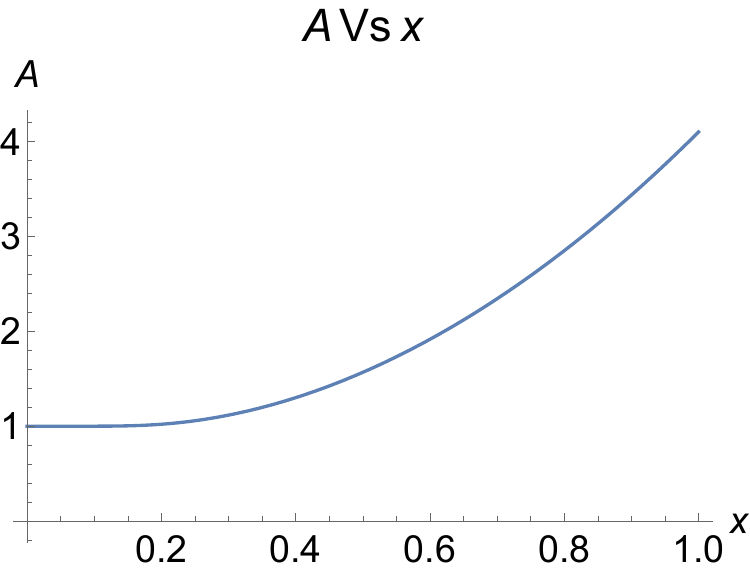}
\caption{$A$ vs $x=\frac{1}{\beta\omega}$ plot: It shows that $A$ increases (i.e. purity decreases) with the increase of temperature, implying that the temperature favors entanglement.}
\label{Plot 1:Test123}
\end{figure}

In general the Purity for a general state, represented by $\rho$, is defined by $P=\textrm{Tr}(\rho^2)$. To find the change of Purity of the first oscillator's thermal vacuum state, we concentrate on (\ref{eq:reduced_density_matrix}), obtained after tracing out \( 2, \tilde{2} \). In this case, neglecting terms of $\mathcal{O}(g^3)$ and higher, we find the Purity at later time as 
\begin{eqnarray}
    P_{\beta}(t) &=& \text{Tr}_{1,\Tilde{1}}\left((\rho_{1,\Tilde{1}}^\beta(t))^2\right) \approx 1 - 4\abs{c_+}^2 \Bigg\{ 1 + \frac{3e^{-\beta\omega}}{(1-e^{-\beta\omega})^2} 
    + \frac{e^{-2\beta\omega}}{(1-e^{-\beta\omega})} + \frac{e^{-3\beta\omega}}{(1-e^{-\beta\omega})^2} \Bigg\}
    \nonumber
    \\
    &=& 1 - 4\abs{c_+}^2 A~,
    \label{BRM1}
\end{eqnarray}
where 
\begin{equation}
A=1+\frac{3e^{-\beta\omega}}{(1-e^{-\beta\omega})^2}+\frac{e^{-2\beta\omega}}{(1-e^{-\beta\omega})}+\frac{e^{-3\beta\omega}}{(1-e^{-\beta\omega})^2}~.
\end{equation}
Note that, in the absence of gravitational waves (i.e., \( c_+ = 0 \)), the purity remains unity, confirming that the initial state of \( \rho_{1,\Tilde{1}}^\beta(t) \) is a pure thermal vacuum state. Since gravitational waves introduce direct coupling between the two physical modes \( 1 \) and \( 2 \), as well as between the two thermal modes \( \tilde{1} \) and \( \tilde{2} \), tracing out \( 2, \tilde{2} \) induces mixedness in the reduced state due to the loss of information about their gravitationally coupled partners. This deviation from purity is quantified by Eq. \eqref{BRM1}. However we see that the temperature alone can not put any effect on the state; it is GW which alone is sufficient to affect the initial state (see with $\beta\to\infty$ limit). So the temperature of the HO is just playing the role of catalyst only in presence of GW. A quantitative feature of purity change with time can be given once a suitable form of the GW is being chosen. This we leave for the next section. A variation with respect to temperature can be done immediately. For that, below we plot $A$ as a function of $x = 1/(\beta\omega)$ in Figure \ref{Plot 1:Test123}. 
It shows that for a given gravitational wave, the purity decreases more at higher temperature. This implies that the entanglement between the HOs is enhanced by the background temperature. However, it must be noted that the above quantifies the entanglement between the thermal vacuum states of the HOs. 

\begin{figure}[htbp]
\centering
\includegraphics[width=0.6\linewidth]{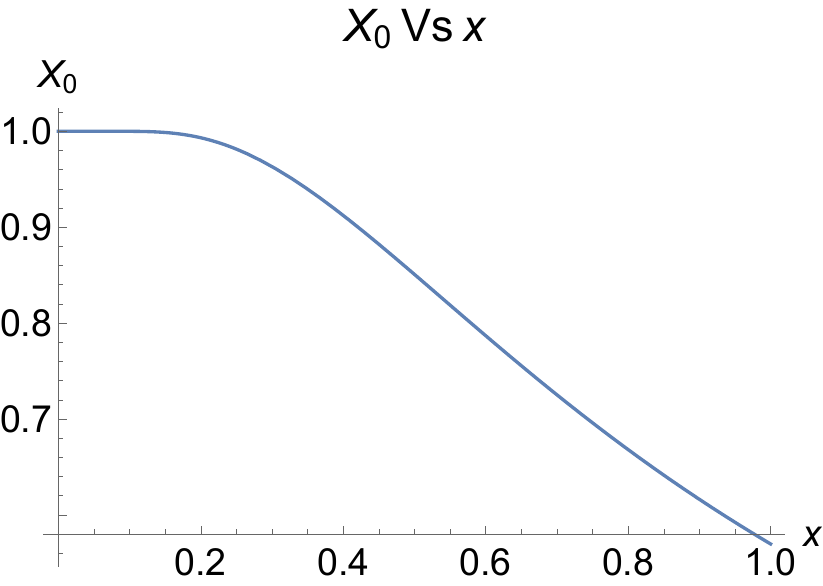}
\caption{Plot of $X_{0}(\beta)$ as a function of $x=1/(\beta\omega)$: Purity for initial state corresponding to first physical HO is unity when $\beta\to\infty$ and decreases with the increase of temperature, showing that for finite $\beta$ the initial state is mixed one.}
\label{fig:X0_beta_plot}
\end{figure}

Furthermore, the entanglement between the fictitious thermal modes $\tilde{1}$ and $\tilde{2}$ due to the gravitational interaction is not directly observable but serves as a mathematical construct that encapsulates how temperature influences gravitationally induced quantum correlations. Since $\tilde{2}$ was already traced out in the previous step, tracing out $\tilde{1}$ removes any explicit manifestation of entanglement between the two thermal modes at the level of the reduced density matrix. However, because $1$ and $\tilde{1}$ are thermally correlated, the first oscillator mode retains an indirect imprint of the prior gravitational entanglement between the thermal modes. This contributes to an additional thermal-induced mixedness in the physical system.
Thus, to distinguish the contributions to mixedness from temperature and gravitational wave effects, we proceed by tracing out $\tilde{1}$, obtaining the further reduced state $\rho_{1}^{\beta}(t)$. The corresponding purity is given by:
\begin{equation}\label{eq:P1_beta}
   P_{1\beta}(t) = \textrm{Tr}_1\left((\rho_1^\beta(t))^2\right) \approx X_0(\beta) - 2\abs{c_+}^2 X_2(\beta)~,
\end{equation}
with
\begin{equation}\label{X0}
   X_0(\beta) = \left( 1 - e^{-\beta \omega} \right) - e^{-2\beta \omega} \tanh\left(\frac{ \beta \omega}{2} \right)~;
\end{equation}
and
\begin{eqnarray}\label{X2}
    &&X_2(\beta) = (1 - e^{-\beta\omega})^2 (1 + e^{-\beta\omega}) 
    \nonumber
    \\
    &-& e^{-2\beta\omega} \Bigg[\frac{e^{\beta\omega}}{(1 - e^{-\beta\omega})} 
    - \frac{2 - e^{-2\beta\omega}}{(1 + e^{-\beta\omega})} 
    - \frac{2e^{-\beta\omega}}{(1 + e^{-\beta\omega})(1 - e^{-2\beta\omega})} 
    - \frac{2}{(1 - e^{-2\beta\omega})^2} \Bigg].
\end{eqnarray}
The first term in (\ref{eq:P1_beta}) corresponds to the initial thermal state. Note that it is less than unity (see Figure \ref{fig:X0_beta_plot}), which clarifies the fact that the initial state represents a thermal state and therefore is a mixed one. 
The thermal correlation between $1$ and $\tilde{1}$ ensures that $1$ retains an imprint of the gravitationally induced entanglement between the thermal modes, contributing to the mixedness of the reduced state. Consequently, the second term in (\ref{eq:P1_beta}) quantifies the modification of purity due to the combined effects of GW and temperature. While GW interactions primarily generate entanglement, temperature enhances their influence.

\begin{figure}[htbp]
\centering
\includegraphics[width=0.6\linewidth]{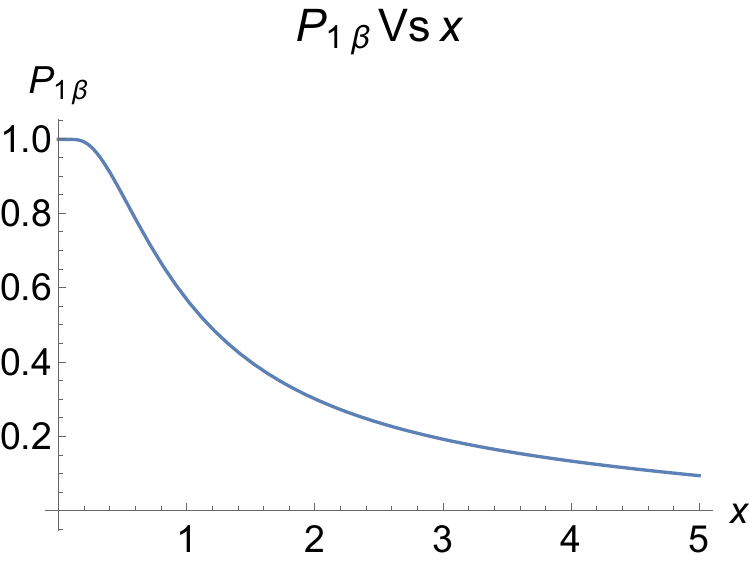}
\caption{Plot of $P_{1\beta}$ as a function of $x=1/(\beta\omega)$ for $|c_+|^2 = 0.001$: Purity of the evolved thermal state decreases with the increase of temperature, implying enhancement of entanglement.}
\label{4}
\end{figure}

Next we plot $P_{1\beta}$ as a function of $x$ in Figure \ref{4} at an instant of time. For this we choose $|c_+|^2 = 0.001$ to ensure the second term of (\ref{eq:P1_beta}) is much less than $X_0(\beta)$ and hence perturbative calculation remains valid.
The figure shows that $P_{1\beta}$ decreases with increasing temperature, indicating enhanced entanglement. This reflects both the direct entanglement between the physical modes induced by gravitational interactions and the residual influence of prior correlations between the thermal modes. Although the thermal modes are not physical, their correlations—before tracing—contribute to the mixedness of the reduced state of $1$, reinforcing the role of temperature as a catalyst in amplifying gravitationally induced entanglement.


Moreover, by taking the limit \( \beta \rightarrow \infty \), Eq. \eqref{eq:P1_beta} aligns with the result obtained in \cite{Nandi:2024zxp}. 
In fact, as shown in Figure \ref{fig:X0_beta_plot}, it is evident that the purity function remains less than unity due to the thermal effects alone. However, in the presence of gravitational wave interactions, the purity deviates further, indicating how the degradation of purity is related to the dynamics of entanglement caused by these interactions. This is clearly demonstrated in the next subsection. 

\vspace{-3mm}


\subsection{Induced entanglement entropy}

In the context of a multimode system, the entanglement between two subsystems induced by gravitational waves can also be characterized by the entropy of one of the reduced states, known as the entanglement entropy. To quantify the entanglement induced by gravitational waves in the thermal states of our quantum system, we will use the quantum Rényi entropy to assess the entanglement of the first thermal harmonic oscillator with other unobserved modes, taking into account both the thermal and gravitational wave effects. The quantum Rényi entropy is a generalization of the von Neumann entropy \cite{Renyi1970} and is defined as
\begin{equation}\label{eq:renyi_entropy}
    S_{\alpha}(\rho) = \frac{1}{1-\alpha} \ln{\mathrm{Tr}(\rho^\alpha)}, \quad \alpha \in (0,1) \cup (1,\infty)~.
\end{equation}
In the limit \(\alpha \to 1\), this reduces to the von Neumann entropy. For \(\alpha = 2\), we obtain the simpler expression for the second-order Rényi entropy:
\begin{equation}\label{eq:renyi_second_order}
S_2(\rho) = -\ln{\mathrm{Tr}(\rho^2)}~.
\end{equation}

Considering terms up to second order in \(g(t)\), the entropy corresponding to the evolved thermal vacuum state of the first HO is given by:
\begin{align}\label{eq:entropy_full_system}
    S_{2\beta}(t) &= -\ln{\mathrm{Tr}_{1,\tilde{1}} \left((\rho_{1,\tilde{1}}^{\beta}(t))^2\right)} \notag \\
    &= -\ln{P_\beta(t)} \notag \\
    &\approx 4\abs{c_+}^2 \left( 1 + \frac{3e^{-\beta\omega}}{(1 - e^{-\beta\omega})^2} 
    + \frac{e^{-2\beta\omega}}{1 - e^{-\beta\omega}} 
    + \frac{e^{-3\beta\omega}}{(1 - e^{-\beta\omega})^2} \right) \notag \\
    &= 4\abs{c_+}^2 A.
\end{align}
Note that, like in purity, the entanglement entropy is generated by GW, whose absence do not produce entanglement between the HOs; however the temperature works as a catalyst to increase it more. Therefore the entanglement between the thermal vacuum states of the two HOs is produced by GWs and their temperature helps to increase this entanglement (see Figure \ref{Plot 1:Test123}).

\begin{figure}[htbp]
\centering
\includegraphics[width=0.6\linewidth]{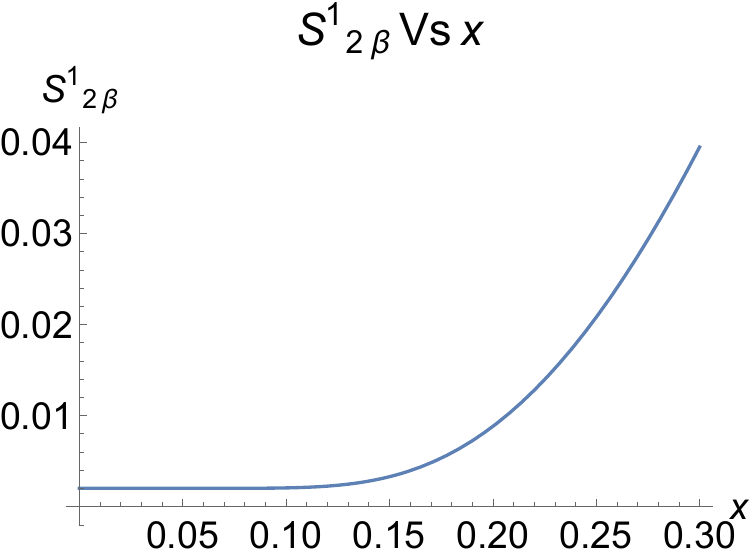}
\caption{Plot of $S^1_{2\beta}$ as a function of $x=1/(\beta\omega)$ for $|c_+|^2 = 0.001$: The entanglement entropy of the evolved thermal state increases with the increase of temperature, showing that the background temperature favors entanglement between two HOs.}
\label{5}
\end{figure}

Similarly, that of the thermal state of HO is the following:
\begin{equation}\label{eq:entropy_physical_system}
    S^1_{2\beta}(t) = -\ln{\mathrm{Tr}_1 \left((\rho_1^\beta(t))^2\right)} = -\ln{P_{1\beta}(t)}~,
\end{equation}
which becomes:
\begin{align}\label{eq:entropy_physical_system_expanded}
    S^1_{2\beta}(t) &= -\ln \Bigg[ (1 - e^{-\beta\omega})^2 \left( 1 - 2\abs{c_+}^2(1 + e^{-\beta\omega}) \right) + e^{-\beta\omega} \left( \frac{1 - e^{-\beta\omega}}{1 + e^{-\beta\omega}} \right) \notag \\
    &\quad + 2\abs{c_+}^2 e^{-2\beta\omega} \Bigg\{ \frac{e^{\beta\omega}}{1 - e^{-\beta\omega}} 
    - \left(\frac{2 - e^{-2\beta\omega}}{1 + e^{-\beta\omega}}\right) \notag \\
    &\quad - \frac{2e^{-\beta\omega}}{(1 + e^{-\beta\omega})(1 - e^{-2\beta\omega})} 
    - \frac{2}{(1 - e^{-2\beta\omega})^2} \Bigg\} \Bigg]~.
\end{align}
This simplifies to 
\begin{equation}
    S^1_{2\beta}(t) \simeq - \ln{X_0(\beta)} + 2|c_+|^2\frac{X_2(\beta)}{X_0(\beta)}~,
    \label{h}
\end{equation}
where \(X_0(\beta)\) and \(X_2(\beta)\) are functions of the inverse temperature \(\beta\), as defined in equations \eqref{X0} and \eqref{X2}. Consequently, like purity, the entropy of the subsystem is also influenced by classical gravitational waves. Moreover, this shows that the presence of gravitational wave induce entanglement between the thermal states of two HOs, modulated by thermal effects. Since $X_0(\beta)$ is less than unity (see Figure \ref{fig:X0_beta_plot}), the first term of the last expression is positive. On the other hand Figure \ref{5}
shows that the entropy increases with the increase of temperature. Therefore this analysis again establishes that the entanglement between the thermal states of the HOs is favored by the background temperature.  

\vspace{-1mm}

Furthermore, it is important to note that the above expression for the entanglement entropy depends on the time-dependent periodic parameter \(\mid c_{+}(t)\mid\), which is influenced by the periodicity of the external coupling parameter \(g(t)\) and the frequency of the system's free oscillatory component. This relationship arises through the time evolution of the states, as described in equation \((\ref{j})\). Consequently, the entanglement entropy exhibits periodic variations, a phenomenon commonly referred to as entanglement dynamics. In the next section, we will discuss more about this entanglement dynamics.

\section{Gravitationally induced time-modulated quantum state}\label{Sec4}
\subsection{Induction of mixing in energy distribution}
As discussed in the previous section, thermal effects alone are insufficient to generate entanglement entropy between two independent physical modes of the oscillator detector. In contrast, interactions with gravitational waves drive the global dynamical evolution of entanglement. Therefore, it is crucial to investigate how gravitational waves locally influence the thermal state of the oscillator modes in the detector.

To explore this, we analyze the energy distribution (in units of \( \hbar\omega \)) for subsystem ``1," which oscillates with frequency \( \omega \), in the thermal state \( \rho^1_{\beta} \). Specifically, this involves calculating the expectation value of the number operator, \( \langle \hat{N}_1 \rangle_{\beta, g} \), where \( \hat{N}_1 = \hat{a}_1^\dagger \hat{a}_1 \), corresponding to the first mode of a 2D oscillator detector in a thermal state. This analysis accounts for the effects of gravitational wave interactions. The oscillator subsystem ``1" can be effectively treated within the framework of the canonical ensemble, as it exchanges energy with the thermal bath in the presence of gravitational wave interactions.

The energy distribution of subsystem 1, expressed in units of \( \hbar\omega \), is given by:  
\begin{equation}
\langle \hat{N}_1 \rangle_{\beta, g} = \text{Tr}_1 \left( \hat{N}_1 \hat{\rho}_1^{\beta}(t) \right).
\end{equation}
By substituting the explicit form of \(\hat{\rho}^1_{\beta}\) from Eq.~(\ref{gh}), we obtain:
\begin{eqnarray}
   \langle \hat{N}_1 \rangle_{\beta,g} &=& \text{Tr}\left( \rho^{\beta}_1(0) \hat{N}_1 \right) 
   + \text{Tr}\left( \delta \rho^{\beta}_{1}(t) \hat{N}_1 \right) 
   \nonumber
   \\
   &=& \sum_{n_{x}=1}^{\infty} \left[ n_{x} A_{n_{x}} + |c_+(t)|^2 n_{x} B_{n_{x}} \right] 
   + |c_+ (t)|^2  \sum_{n_{x}=0}^{\infty} (n_{x}+1) C_{n_{x}} 
   \nonumber
   \\
   &=& \frac{1}{e^{\beta \omega} - 1} 
   + |c_+ (t)|^2 \coth(\frac{\beta\omega}{2}).
   \label{hg}
\end{eqnarray}
In the absence of gravitational interactions (\(g = 0\)), the energy distribution simplifies to the standard thermal result:
\begin{equation}
\langle \hat{N}_1 \rangle_{\beta, g=0} = \langle \hat{N}_1 \rangle_{\beta} = \frac{1}{e^{\beta \omega} - 1}.
\end{equation}
On the other hand, at zero temperature (\(\beta \to \infty\)), the expectation value of \(\hat{N}_1\) becomes:
\begin{equation}
\langle \hat{N}_1 \rangle_{\beta \to \infty, g} = \langle \hat{N}_1 \rangle_{g} = |c_{+}(t)|^2,
\end{equation}

Notably, in the zero-temperature limit, the reduced density matrix describing the physical subsystem in Eq. (\ref{gh}) bears a structural resemblance to the leading-order approximation of a quantum harmonic oscillator interacting with an effective thermal bath (see, for instance, \cite{Pal:2020jvq}), leading to a thermal mixed state. Specifically, when truncated to the first excited state, the thermal density matrix takes the form:  
\begin{equation}
\rho_{th} = (1 - e^{-\frac{\omega}{k_{B}T_{\text{eff}}}}) \sum_{n_1=0}^{\infty} e^{-\frac{n_1 \omega}{k_{B}T_{\text{eff}}}} \ket{n_1}\bra{n_1} \approx (1 - e^{-\frac{\omega}{k_{B}T_{\text{eff}}}}) \ket{0}\bra{0} + e^{-\frac{\omega}{k_{B}T_{\text{eff}}}} \ket{1}\bra{1},
\label{z}
\end{equation}  
where the approximation \( e^{-\frac{\omega}{k_{B}T_{\text{eff}}}} \ll 1 \) holds. The resemblance between Eq. (\ref{gh}) and the mixed state in Eq. (\ref{z}) allows us to identify an effective thermal characterization of the system, leading to the definition of a thermal Maxwell-Boltzmann-type distribution: 
\begin{equation}  
|c_{+}|^2 = \exp\left(-\frac{\omega}{k_B T_{\text{eff}}(t)}\right).  
\end{equation}  
From this, we define the effective temperature \(T_{\text{eff}}(t)\):
\begin{equation}  
T_{\text{eff}}(t) = -\frac{\omega}{k_B \ln\left(|c_+(t)|^2\right)}.
\label{t}
\end{equation} 
Note that the above temperature is time dependent and hence it is a non-equilibrium situation. In this case the thermodynamic quantities are not defined properly, as done in equilibrium cases. However, there are suggestions (although not fully understood) to interpret the thermodynamic quantities in the non-equilibrium situation as well as identifying these quantities (e.g. see \cite{J2003}). Our present temperature has been identified following these existing ideas.  

Thus, the energy distribution in (\ref{hg}) can be expressed as contributions from both the thermal distribution and the interaction with gravitational waves, as follows:
\begin{equation}
\langle \hat{N}_1 \rangle_{\beta, g} = \langle \hat{N}_1 \rangle_{\beta} + \langle \hat{N}_1 \rangle_{g} + 2 \langle \hat{N}_1 \rangle_{\beta} \langle \hat{N}_1 \rangle_{g}~.
\label{k}
\end{equation}
The above results shows that the GW itself provides an independent thermalization in the system, represented by term $\langle \hat{N}_1 \rangle_{g}$. We call this as {\it induced thermalization} of the system. On the other hand $\langle \hat{N}_1 \rangle_{\beta}$ is the usual term as the initial state is a thermal state for the first HO, which can be termed as {\it spontaneous thermalization}. However, the last term is very interesting. It will be present when both the background thermality and the interaction with GW are existing. Therefore, it can be interpreted as the {\it stimulated thermalization}. A similar structure was also noticed earlier in the excitation probability of a two-level atom which is interacting with the background thermal fields \cite{Kolekar:2013xua,Kolekar:2013hra,Chowdhury:2021ieg,Ghosh:2024mqy} (also see \cite{Chowdhury:2019set,Barman:2021oum,Barman:2023bpz}).
Furthermore, this result highlights a hybrid statistical behavior in which the mean occupation number reflects contributions from both quantum (Bose-Einstein) and classical (Maxwell-Boltzmann) distributions. A similar phenomenon has been explored in statistical models where a ``transmutational potential" facilitates transitions between particle states, leading to mixed statistical distributions \cite{PhysRevE.49.5111}. In the present context, gravitational wave interactions play an analogous role, acting as a physical mechanism that mediates transitions between quantum and classical regimes. Specifically, when gravitational waves interact with a thermal equilibrium system, they perturb it, driving the system into a non-equilibrium state. This transition causes the statistical distribution to deviate from standard equilibrium forms such as the Bose-Einstein distribution, commonly associated with quantum oscillators. Instead, the system adopts a time-dependent, non-equilibrium distribution that evolves dynamically under the influence of gravitational wave interactions, reflecting the departure of the system from its initial equilibrium configuration.

\subsection{A memory-retaining `time-crystal' phase}

A particularly intriguing feature of our system is the periodic dynamics of entanglement entropy and the statistical distribution at finite temperature, induced by the presence of gravitational waves. In this context, the reduced density matrix of subsystem 1, denoted as $\rho_1^\beta(t)$, deviates from thermal equilibrium due to time-dependent corrections. Specifically, these corrections, $\delta \rho^{\beta}_g(t)$, to the equilibrium distribution $\rho^{\beta}_0$ (see Eq. \((\ref{k})\)), are explicitly time-dependent and take the form:
\begin{equation}
\delta \rho^{\beta}_g(t) \sim |c_+(t)|^2,
\end{equation}
where $|c_+(t)|^2$ is a periodic function of time.

It is important to note that the full system's Hamiltonian (\ref{H2}) satisfies the periodicity condition imposed by the gravitational wave-driven perturbation (namely $g(t)$):
\begin{equation}
H(t+T_g) = H(t), \quad T_g = \frac{2\pi}{\omega_g},
\end{equation}
where $\omega_g$ is the gravitational wave frequency.
For instance, in case of a linearly polarized gravitational wave, the wave parameters are given by \cite{Nandi:2024zxp}
\begin{eqnarray}
&&\gamma(t) = -\frac{1}{2} \omega_g \chi_0 \epsilon_+ \sin(\omega_g t),
\\
&&\delta(t) = -\omega_g \chi_0 \epsilon_\times \sin(\omega_g t).
\end{eqnarray}
The above ones can be realized as follows. For the choice of GW as $\chi(t) = \chi_0 \cos(\omega_g t)$ ($\chi(t)$ has appeared in terms given below Eq. (\ref{A.4})), the identification of $\gamma$ and $\delta$, given in Appendix \ref{App1} leads to the above expressions. 
Substituting these into Eq. \((\ref{j})\), the squared magnitude of the time-dependent coefficient $|c_+(t)|^2$ is obtained as:
\begin{eqnarray}
|c_+(t)|^2 &=& \left(\frac{\chi_0 f}{1-f^2}\right)^2 \bigg[(1+f^2) - (1-f^2)\cos^2(\omega_g t)  
- 2f^2\cos(\omega_g t)\cos(\Omega t) 
\nonumber
\\
&-& 2f \sin(\omega_g t)\sin(\Omega t)\bigg],
\label{me}
\end{eqnarray}
where $f = \frac{\omega_g}{\Omega}$ is the ratio of the gravitational wave frequency to double of the natural frequency of the system.

This result clearly demonstrates that the reduced density matrix $\rho^\beta_1(t)$ does not strictly follow the periodicity of the external gravitational wave driving. Specifically, we find:
\begin{equation}
\rho^\beta_1(t + T_g) \neq \rho^{\beta}_1(t), \quad T_g = \frac{2\pi}{\omega_g}.
\end{equation}

\begin{figure}[htbp]
\centering
\includegraphics[width=0.9\linewidth]{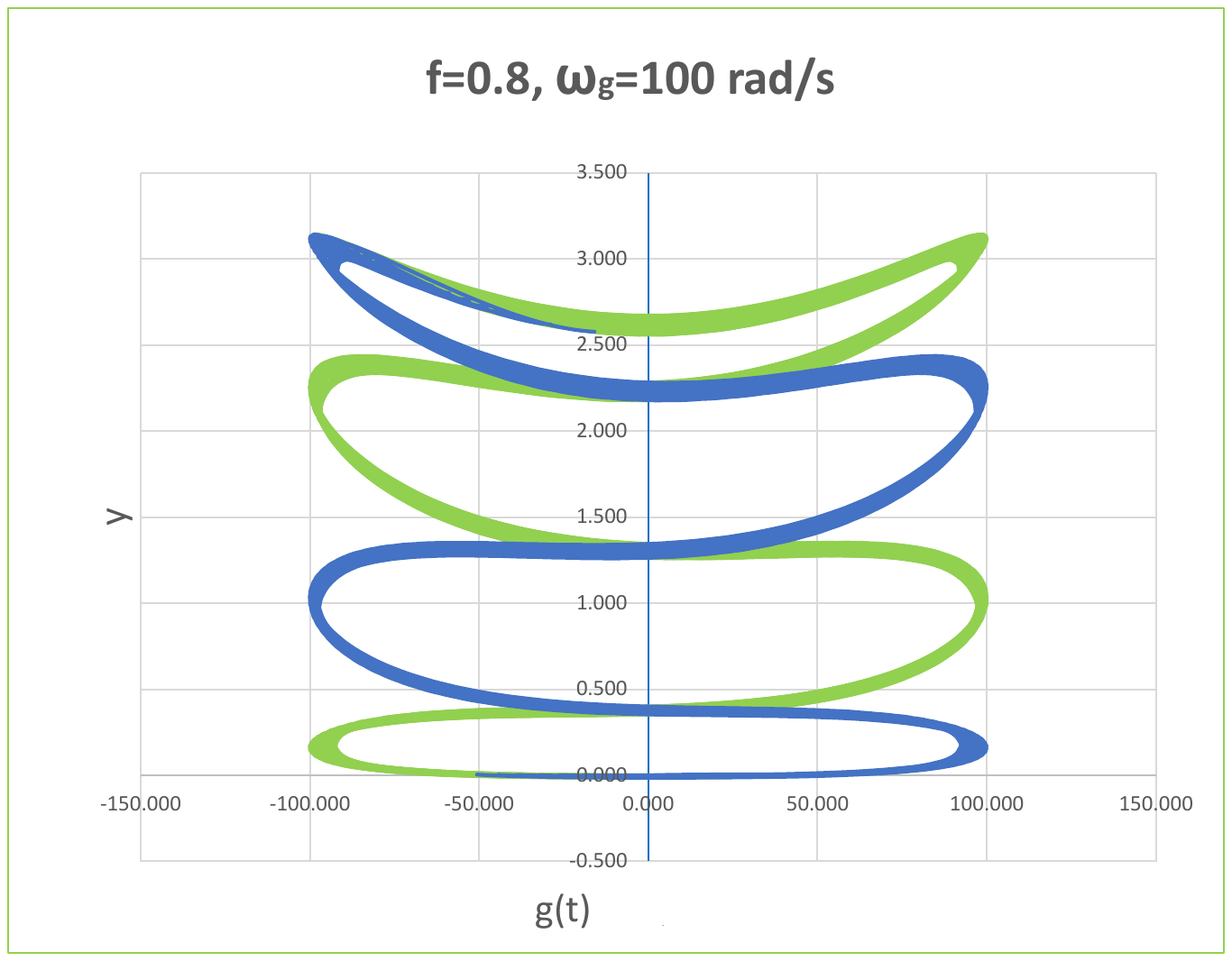}
\caption{$y$ vs $g(t)$ curve for $f=0.8$ and $\omega_g=100$ rad/sec: The effect of GWs (quantified by $y$) in the purity and entropy does not vanish for a span of time, although GWs (i.e. $g$) vanishes several times within this span, implying retention of memory of GWs within the quantum state.}
\label{memory}
\end{figure}

This feature introduces a novel aspect that corroborates the quantum memory effect induced by gravitational waves, as follows:
\begin{itemize}
\item It is noteworthy that when the mechanical frequency scale of the subsystem, $\Omega$, and the gravitational wave frequency $\omega_g$ are commensurate (i.e., their ratio is a rational number), the subsystem oscillation in Eq. (\ref{me}) becomes periodic, with a period determined by the least common multiple (LCM) of their respective periods. Consequently, the system will eventually return to its initial state. However, in the incommensurate case (when their ratio is an irrational number), the system does not settle into a strictly periodic state but instead exhibits quasi-periodic or modulated behavior.

\item In the commensurate case, while the Hamiltonian of the full system exhibits strict periodicity, the reduced subsystem undergoes oscillations with a distinct periodicity, forming a subharmonic response relative to the gravitational wave period. This indicates an emergent symmetry-breaking phenomenon within the quantum subsystem, which contrasts with the externally driven periodicity imposed by the gravitational waves.

\item As we have noted, in the zero-temperature limit, when the gravitational wave (GW) interaction periodically vanishes, the subsystem does not oscillate exactly in sync with the external drive. Instead, it oscillates at a different frequency—often larger than the driving frequency. Moreover, due to induced thermalization, the subsystem remains in an effectively low-temperature thermal state, with the effective temperature itself being a periodic function of time, as shown in Eq.~(\ref{t}). This suggests that the system maintains instantaneous thermal equilibrium rather than global equilibrium and does not immediately revert to a pure state. As a result, a persistent dynamical imprint of the gravitational influence emerges, leading to the development of a quantum memory effect, where past interactions with gravitational waves continue to influence the subsystem's state.

This behavior is illustrated in Figure \ref{memory}, where we plot $y= |c_+(t)|^2/ \left(\frac{\chi_0 f}{1-f^2}\right)^2$ as a function of the GW wave strength $g(t)$. Initially, at $g(t=0)=0$, we have $y=0$. However, as time progresses, $y$ increases, even when $g(t)$ fluctuates between zero and a maximum value (see the blue region of the curve). After some time, $y$ starts decreasing and eventually returns to its initial value of zero (see the green region of the curve). Notably, in the intermediate stages, even when the interaction vanishes, $y$ does not immediately vanish. Depending on the choice of parameters, it remains nonzero for a sufficiently long time. This confirms that the quantum state of the harmonic oscillator retains memory of past interactions, a phenomenon we refer to as the ``quantum memory effect'' induced by gravitational waves. This dynamical nature is reminiscent of the hysteresis observed in magnetic materials subjected to a variable magnetic field.

\item Furthermore, at finite background temperature, an exotic mixing of statistical distributions emerges (\ref{k}), indicating that the subsystem deviates from equilibrium, going beyond the standard Bose-Einstein thermal energy distribution. However, thermal effects do not immediately destroy the quantum memory effect induced by gravitational waves. Instead, the system retains information about its past interactions with gravitational waves and does not instantly return to thermal equilibrium. This suggests a new form of prethermalization, arising from the interplay between periodically induced and spontaneous thermalization, both governed by gravitational influence. When the memory effect persists but gradually diminishes over a time scale much longer than the period of the driven gravitational interaction, the system slowly transits toward thermal equilibrium.

Moreover, at finite temperature, this quantum memory effect can be interpreted as a gravitationally induced time-crystal-like phase under periodic driving, resembling prethermal time crystals (PTCs), which were recently explored in \cite{PhysRevX.7.011026} and serve as nonequilibrium extensions of Floquet time crystals \cite{PhysRevLett.117.090402}. These systems spontaneously break discrete time-translation symmetry and exhibit a periodicity distinct from that of the external drive. However, unlike standard PTCs, which eventually thermalize due to heating from the periodic drive, our system resists full thermalization as long as the memory effect persists. In our case, the gravitationally induced memory effect at finite temperature shares key features with prethermal time crystals rather than representing an exact PTC. We refer to this as a \textit{prethermal time-crystal-like phase}. Rather than being directly erased by temperature, the memory effect gradually fades over time. Once it fully vanishes at a specific moment, the system naturally relaxes back to thermal equilibrium, with the environmental temperature continuously shaping this transition.
\end{itemize}

The concept of quantum time crystals (QTC), originally proposed by Wilczek and Shapere as systems that spontaneously break continuous time-translation symmetry, follows a fundamentally different mechanism \cite{PhysRevLett.109.160401, Sacha:2017fqe}. However, in our case, the deviation of the subsystem does not result from spontaneous symmetry breaking but instead arises from the external periodic influence of gravitational waves. Since the subsystem’s periodicity is externally imposed and constrained by the oscillator frequency, its oscillatory behavior constitutes a forced response rather than spontaneous symmetry breaking.
Moreover, it has been recently shown in \cite{bernardini2022emergent} that time crystal phases \cite{PhysRevLett.109.160401} can arise without explicitly requiring spontaneous breaking of time translation symmetry. This phenomenon has been observed in driven quantum systems where periodic behavior emerges through alternative mechanisms, such as noncommutativity in phase space \cite{Bertolami:2024tom}. In light of these findings, our results further support the idea that gravitational interactions can induce similar oscillatory phases, albeit with key distinctions, as outlined in Table~\ref{table:comparison}. 
\begin{table}[h!]
\centering
\renewcommand{\arraystretch}{1.3}
\setlength{\tabcolsep}{8pt}
\begin{tabular}{|p{4cm}|p{5cm}|p{5cm}|}
\hline
\textbf{Feature} & \textbf{Prethermal time crystal} & \textbf{Our case} \\
\hline
\textbf{Origin of periodicity} & Spontaneous subharmonic selection & Externally driven by gravitational waves \\
\hline
\textbf{Time translation symmetry} & Spontaneously broken & Imposed by external drive \\
\hline
\textbf{Driving mechanism} & Floquet periodic drive & Gravitational wave interaction \\
\hline
\textbf{Frequency selection} & Independent subharmonic & Matches imposed drive frequency \\
\hline
\textbf{Does it retain memory across cycles?} & Yes, retains periodicity before thermalization & Yes, remains out of equilibrium even during ``off" phases \\
\hline
\end{tabular}
\caption{Comparison between prethermal time crystals and our system}
\label{table:comparison}
\end{table}
This reinforces the notion that time crystal-like behavior can manifest under broader physical conditions, including systems where periodic responses are externally driven rather than self-organized.

\section{Conclusion and Outlook}\label{Sec5}

In this study, we investigated the interplay between gravitational waves and thermality in a system of quantum harmonic oscillators, modeling the endpoints of interferometer arms in LIGO-like detectors as two-dimensional oscillators. Using the thermofield dynamics (TFD) formalism, we demonstrated that gravitational waves induce entanglement between initially unentangled thermal oscillators. However, thermal effects alone cannot generate entanglement between two independent physical modes; instead, thermality modifies the entanglement dynamics induced by gravitational waves. This distinction highlights two fundamental aspects of our system: (i) the role of thermality in altering statistical properties and (ii) the role of gravitational waves in producing a quantum memory effect. 

A key result of our study is the identification of a gravitationally induced discrete time-translation symmetry breaking in the reduced subsystem, leading to a persistent quantum memory effect. While the full Hamiltonian remains strictly periodic under the external gravitational wave drive, the reduced subsystem oscillates with a distinct periodicity, forming a subharmonic response. However, this is not a spontaneous symmetry-breaking phenomenon but an externally imposed symmetry breaking arising from the interplay between the natural oscillator frequency and the frequency of the gravitational wave drive. As a consequence, even when the gravitational wave interaction periodically vanishes, the subsystem does not immediately relax to equilibrium but retains a dynamical imprint of past interactions, remaining out of equilibrium despite the absence of an external force.

Furthermore, at finite temperature, an exotic mixing of statistical distributions emerges, causing the subsystem to deviate from standard Bose-Einstein thermal equilibrium. However, thermal effects do not destroy the quantum memory effect; rather, they influence the system’s overall evolution without altering the intrinsic periodic nature of the memory effect itself. Even in the complete absence of temperature, the quantum memory effect persists, governed solely by gravitational wave interactions.

The analogy with prethermal time crystals (PTCs) is meaningful only in the presence of a background temperature, where thermality modifies the behavior of the quantum memory effect. While PTCs exhibit spontaneous discrete time-translation symmetry breaking and eventually thermalize due to heating from a periodic drive, our system follows a distinct trajectory. The reduced subsystem undergoes externally driven discrete time-translation symmetry breaking rather than spontaneous symmetry breaking. Moreover, as long as the memory effect persists, the system resists full thermalization. Once the memory effect gradually diminishes over a timescale much longer than the gravitational driving period, the system naturally evolves back to thermal equilibrium.

Overall, our findings suggest that gravitational wave interactions can give rise to a novel prethermal time-crystal-like phase, but only in the presence of a background temperature. This phase is characterized by externally imposed discrete time-translation symmetry breaking and a memory effect that persists across multiple cycles of the system’s evolution. Unlike conventional prethermal time crystals, where thermalization is governed by slow heating effects, our system maintains nontrivial correlations through the quantum memory effect. This work provides new insights into gravitationally induced quantum correlations in thermal environments, opening potential avenues for further studies on the interplay between quantum memory, thermalization, and gravitational interactions in high-precision interferometric setups.

{\it {Broader Conceptual Structure and Physical Implications. --}}
Taken together, these findings motivate a closer look at the foundational structure of our model. Beyond the dynamical results, the interplay between geometry, entanglement, and thermality points toward a broader theoretical framework, which we now summarize:

\begin{itemize}
 \item[(i)] While it is technically possible to perform the analysis using the standard density matrix formalism—by evolving a mixed thermal state (represented by thermal density matrix) under the time-dependent Hamiltonian—this approach has significant limitations. It obscures the entanglement structure between the physical system and the thermal sector, making it challenging to disentangle the contributions of gravitationally induced coherence from thermal noise. Furthermore, the repeated need for partial traces and manipulations of mixed states in such a framework can be analytically intractable in the presence of time-dependent squeezing interactions. In contrast, the TFD formalism keeps both sectors explicit, preserves coherence throughout, and simplifies the treatment of entanglement.
\item[(ii)] The TFD approach maintains unitary evolution in an extended Hilbert space, treating thermal states as pure states which are constructed by entangling the physical modes with the auxiliary (tilde) modes. This representation retains full access to quantum correlations throughout the evolution and enables operator-level tracking of entanglement across both physical and thermal sectors.
\item[(iii)] The framework thus provides a powerful and elegant way to algebraically isolate gravitationally induced entanglement from its thermal modifications. This makes possible exact computation of observables like purity, entropy, and effective temperature—quantities otherwise can be inaccessible through conventional mixed-state treatments.
\item[(iv)] The interaction part of the Hamiltonian is derived directly from the geodesic deviation equation (as detailed in Appendix \ref{Apprev1}) and exhibits a momentum--position coupling structure, in contrast to the more conventional position--position or dipole-type interactions. This form encapsulates the geometric influence of spacetime curvature on the quantum phase space. Importantly, the resulting quadratic Hamiltonian terms, which include pair creation and annihilation operators, generate an algebra which is closed under the non-compact group \( \mathrm{SU}(1,1) \).
This \( \mathrm{SU}(1,1) \) algebraic structure is particularly relevant in our case because it governs the dynamics of two-mode squeezing, which is the key mechanism by which the gravitational wave induces entanglement between the oscillators. The algebraic closure enables exact diagonalization and analytic time evolution, allowing us to compute observables like purity and entropy in closed form. Thus, the symmetry structure not only reflects the physical nature of the coupling but also renders the system exactly solvable within the TFD framework.
\item[(v)] The memory effect we observe is a coherent dynamical signature of this gravitationally induced entanglement. It manifests as residual modulations in purity and entropy, even after the external driving subsides—capturing an integrated imprint of the system's prior gravitational interaction history.
Importantly, this memory is both \emph{nonlocal} and \emph{global} in the quantum-information-theoretic sense:
    \begin{itemize}
        \item It is \emph{nonlocal} because it arises from entanglement between physically distinct modes; it cannot be attributed to the dynamics of any single oscillator in isolation.
        \item It is \emph{global} because the reduced state of a subsystem reflects the \emph{accumulated interaction history}—not merely the instantaneous value of the gravitational field.
    \end{itemize}
\item[(vi)] Unlike dissipative systems where decoherence originates from energy exchange with an environment, our reduced dynamics are non-unitary due to entanglement with the rest of the system. The memory is thus quantum-informational, not thermodynamic, in nature.
\item[(vii)] The correlations responsible for this effect encode information inaccessible to any local observable alone. Observables like subsystem purity or entropy serve as indirect probes of this global structure.
\item[(viii)] While similar signatures—such as residual coherence and time-dependent purity oscillations—can arise in engineered systems with squeezing interactions, non-Markovian baths, or feedback control (e.g., in cavity QED or optomechanics), our model differs in two key respects. First, the coupling in our system is not artificially designed but emerges naturally from spacetime curvature in linearized gravity. Second, the thermal environment is not an abstract decoherence model but is treated explicitly and coherently via the Thermo Field Dynamics formalism.
\end{itemize}

These conceptual clarifications enhance the interpretative context of our results and may invoke future explorations of gravitationally induced quantum phenomena, particularly in the presence of finite-temperature environments.

\section*{Acknowledgements}
MD acknowledges funding support for the Junior Research Fellowship (JRF) from Indian Institute of Technology Guwahati, India. PN acknowledges the support of the National Institute for Theoretical and Computational Sciences (NITheCS) through the Rector’s Postdoctoral Fellowship Program (RPFP). He also thanks Prof. Frederik G. Scholtz, Prof. Nigel Bishop, and NITheCS Director Prof. Francesco Petruccione for their insightful discussions and valuable academic support. BRM thanks the people of India for supporting basic sciences.

\begin{appendix}
\section*{Appendices}

\section{Derivation of the effective interaction Hamiltonian from geodesic deviation}\label{Apprev1}

To derive the interaction Hamiltonian presented in Eq.~(\ref{x1}) of the main text, we begin with the relativistic geodesic deviation equation. This equation describes the relative acceleration between neighboring geodesics in curved spacetime. Unlike the motion of a single test particle—which can be locally trivialized via the equivalence principle—the relative separation between nearby trajectories is a physically meaningful and gauge-invariant quantity.
The covariant form of the geodesic deviation equation is
\begin{equation}
\frac{D^2 \xi^\mu}{D \tau^2} = - R^\mu_{\ \nu\rho\sigma} \, \xi^\rho \, \frac{dq^\nu}{d\tau} \, \frac{dq^\sigma}{d\tau},
\label{eq:geodesic_dev_covariant}
\end{equation}
where \( \xi^\mu \) denotes the infinitesimal separation vector between neighboring geodesics  and \( D/D\tau \) is the covariant derivative along the reference trajectory. $q^\mu$ represents the event point on the spacetime and $\tau$ is the proper time.

We consider a congruence of timelike geodesics \( q^{\mu}(\tau) \), centered around the reference worldline of a freely falling detector. The analysis is performed in the \emph{proper detector frame}, a comoving coordinate system constructed along the detector's trajectory. In this frame, the metric is locally Minkowskian at the origin, but tidal effects due to spacetime curvature manifest at leading order in spatial separation. The four-velocity of the detector is \( u^\mu = (1, 0, 0, 0) \), and we take the separation vector \( \xi^\mu \) to be spacelike and orthogonal to the reference geodesic, satisfying \( \xi^\mu u_\mu = 0 \). This ensures that \( \xi^\mu \) represents the physical spatial displacement between nearby geodesics, as measured by a local observer attached to the detector.
Under these assumptions, only the spatial components of the covariant geodesic deviation equation contribute nontrivially, reducing it to
\begin{equation}
\frac{d^2 \xi^j}{dt^2} = - R^j_{\ 0k0}(t, \vec{r})\, \xi^k.
\label{eq:geodesic_dev_nonrel}
\end{equation}

To relate this equation to gravitational waves, we adopt the framework of linearized gravity, expanding the metric as \( g_{\mu\nu} = \eta_{\mu\nu} + h_{\mu\nu} \) with \( |h_{\mu\nu}| \ll 1 \). The linearized Riemann tensor is
\begin{equation}
R^\mu_{\ \rho\sigma\nu} = \frac{1}{2} \eta^{\mu\lambda} \left( \partial_\sigma \partial_\rho h_{\nu\lambda} - \partial_\sigma \partial_\lambda h_{\nu\rho} - \partial_\nu \partial_\rho h_{\sigma\lambda} + \partial_\nu \partial_\lambda h_{\sigma\rho} \right).
\end{equation}
Although the metric perturbation \( h_{\mu\nu} \) is gauge-dependent, transforming under infinitesimal coordinate shifts as \( h_{\mu\nu} \rightarrow h_{\mu\nu} + \partial_\mu \epsilon_\nu + \partial_\nu \epsilon_\mu \), the Riemann tensor \( R^\mu_{\ \rho\sigma\nu} \) is gauge invariant.
To isolate the physical degrees of freedom, we impose the transverse-traceless (TT) gauge:
\begin{equation}
h_{0\mu} = 0, \quad \partial^i h_{ij} = 0, \quad h^i_{\ i} = 0.
\end{equation}
In this gauge, the relevant curvature component becomes
\begin{equation}
R^j_{\ 0k0}(t,\vec{r}) = - \frac{\partial \Gamma^j_{\ 0k}}{\partial t}.
\end{equation}
To further simplify, we adopt the \emph{long-wavelength approximation}, which is valid when the gravitational wavelength greatly exceeds the spatial extent of the detector. In this regime, spatial variations of the gravitational wave across the detector are negligible, and the metric perturbation can be treated as purely time-dependent:
$h_{ij}(t, \vec{r}) \approx h_{ij}(t)$.
The geodesic deviation equation then reduces to a purely time-dependent form:
\begin{equation}
\frac{d^2 \xi^j}{dt^2} = - R^j_{\ 0k0}(t)\, \xi^k.
\label{eq:geodesic_dev_nonrel_longwave}
\end{equation}
This is the nonrelativistic limit of the geodesic deviation equation in a temporally varying gravitational wave perturbation, 
where in the Christoffel symbols are given by \( \Gamma^j_{\ 0k} = \frac{1}{2} \dot{h}^j_{\ k} \).

The equation (\ref{eq:geodesic_dev_nonrel_longwave}) can be viewed as two-dimensional non-relativistic motion of a particle with respect to the frame of the other particle under a force given by the right hand side. If there are other external forces except that influenced by the GWs, those can be included easily in the non-relativistic framework. Therefore we now include an external mechanical potential \( V(x) \), leading to the modified equation:
\begin{equation}
m \ddot{x}^j(t) =  m\, \frac{\partial \Gamma^j_{\ 0k}(t)}{\partial t} x^k(t) - \frac{\partial V}{\partial x^j}. \label{eq:geo-dev-eom}
\end{equation}
Our goal is to \textit{reverse engineer} a Lagrangian such that its Euler--Lagrange equation reproduces Eq.~\eqref{eq:geo-dev-eom}. The corresponding Euler--Lagrange operator is given by
\begin{equation}
\mathcal{L}_i(t) = \frac{d}{dt} \left( \frac{\partial L_{\text{Detector}}}{\partial \dot{x}^i} \right) - \frac{\partial L_{\text{Detector}}}{\partial x^i}, \qquad \text{with} \quad \delta S_{\text{Detector}}[x(t)] = -\int dt\, \delta x^i(t)\, \mathcal{L}_i(t). \label{eq:euler-deriv}
\end{equation}
Here we denote $\delta L_{\text{Detector}} = \delta x^i(t)\, \mathcal{L}_i(t)$.
Matching the structure of Eq.~\eqref{eq:geo-dev-eom}, we identify the terms contributing to the Euler derivative:
\begin{equation}
\mathcal{L}_i(t) = m \ddot{x}^i -  m \frac{\partial \Gamma^i_{\ 0k}(t)}{\partial t} x^k + \frac{\partial V}{\partial x^i}.
\end{equation}
Now the variational derivative of \( S_{\text{Detector}} \) with respect to \( x^i \) yields the desired equation of motion~(\ref{eq:geo-dev-eom}). Upon discarding total derivative (surface) terms, we infer that the Lagrangian can be reconstructed by integrating the resulting terms in reverse:
\begin{align}
\delta S_{\text{Detector}}[x(t)] 
&= -\int dt\, \delta x^i(t)\, \mathcal{L}_i(t) \nonumber \\
&= \int dt\, \delta L_{\text{Detector}} \nonumber \\
&= \int dt\, \delta \left[ \frac{1}{2} m \dot{x}^{j\,2}  
     - m \Gamma^j_{\ 0k}(t)\, \dot{x}_j x^k 
     - V(x^j) \right],
\label{eq:euler-deriv}
\end{align}
which allows us to identify the Lagrangian governing the detector’s motion in a weak gravitational field as
\begin{equation}
    L_{\text{Detector}}(x, \dot{x}, t) := L = \frac{1}{2} m \dot{x}^{j\,2} - m \Gamma^j_{\ 0k}(t)\, \dot{x}_j x^k - V(x^j).
\end{equation}

To obtain the Hamiltonian, we compute the canonical momentum:
\begin{align}
p_j &= \frac{\partial L}{\partial \dot{x}^j} = m \dot{x}_j - m \Gamma^j_{\ 0k}(t)\, x^k, \\
\Rightarrow \dot{x}_j &= \frac{1}{m} \left( p_j + m \Gamma^j_{\ 0k}(t)\, x^k \right).
\end{align}
Substituting into the Legendre transformation,
\begin{equation}
H(t) = p_j \dot{x}^j - L,
\end{equation}
we find
\begin{equation}
H(t) = \frac{1}{2m} \left( p_j + m \Gamma^j_{\ 0k}(t)\, x^k \right)^2 + V(x^{j}).
\end{equation}
In the weak-field limit, keeping only linear terms in \( \Gamma \), this reduces to:
\begin{equation}
H(t) = \frac{p_j^2}{2m} + \Gamma^j_{\ 0k}(t)\, x^k p_j + V(x^{j}) + \mathcal{O}(\Gamma^2).
\label{Rev1}
\end{equation}
Furthermore, to advance our analysis, we consider the specific scenario in which the gravitational wave propagates along the \( z \)-axis. In the transverse-traceless (TT) gauge, the transversality condition guarantees that the perturbation \( h_{jk} \) is restricted to the plane orthogonal to the direction of propagation. As a result, the only nonzero components of the metric perturbation reside in the \( x_{1} \)-\( x_{2} \) plane.
In this gauge, the general form of the metric perturbation is given by
\begin{equation}
h_{jk}(t) = 2\chi(t) \left( \epsilon_\times \sigma_{1\,jk} + \epsilon_+ \sigma_{3\,jk} \right),
\label{A.11}
\end{equation}
where \( \chi(t) \) denotes the time-dependent amplitude of the gravitational wave, and \( \epsilon_+ \), \( \epsilon_\times \) are polarization coefficients corresponding to the two physical polarizations: the \emph{plus} and \emph{cross} modes. The symmetric polarization tensors \( \sigma_{1\,jk} \) and \( \sigma_{3\,jk} \) encode the spatial deformation structure within the transverse plane.
Equation~\eqref{A.11} manifests that gravitational waves in the TT gauge affect only the \( x_{1} \)-\( x_{2} \) plane—i.e., the plane orthogonal to the direction of wave propagation. This observation justifies restricting the particle's motion to an effective two-dimensional dynamics for our analysis. A detailed discussion of the polarization modes and their role in entanglement generation is provided in Appendix \ref{App1}.

We now introduce an external mechanical potential modeled as a two-dimensional isotropic harmonic oscillator:
\[
V(x) = \frac{1}{2} m \omega^2 (x_1^2 + x_2^2).
\]
Substituting this into the expression for the Hamiltonian (\ref{Rev1}) derived earlier, the complete Hamiltonian governing the system becomes
\begin{equation}
H(t) = \sum_{j=1}^2 \left( \frac{p_j^2}{2m} + \sum_{k=1}^2 \Gamma^j_{\ 0k}(t)\, x^k p_j + \frac{1}{2} m \omega^2 x_j^2 \right).
\label{H1}
\end{equation}
This Hamiltonian describes a two-dimensional quantum harmonic oscillator subject to velocity–position coupling induced by spacetime curvature—specifically, due to the presence of a time-varying gravitational wave background. It provides the starting point for the quantized treatment developed in the main text.
Upon quantization, the canonical variables \( x^k \) and \( p_j \) are promoted to operators. To ensure hermiticity of the interaction term, we adopt a symmetric ordering:
\begin{equation}
\hat{H}_{\text{int}}(t) = \frac{1}{2} \Gamma^j_{\ 0k}(t)\, \left( \hat{x}^k \hat{p}_j + \hat{p}_j \hat{x}^k \right).
\end{equation}
This completes the derivation of the curvature-induced interaction Hamiltonian from a well-defined action principle based on the geodesic deviation framework.

\section{Derivation of Eq. (\ref{H2})} \label{App1}
In transverse traceless gauge condition, GWs $h_{jk}$ can be chosen as
\begin{equation}\label{A.1}
    h_{jk}=2\chi(t)(\epsilon_\times\sigma_{1jk}+\epsilon_+\sigma_{3jk})~,
\end{equation}
which leads to
\begin{equation}\label{A.2}
    \Gamma^j_{0k}=\frac{1}{2}\partial_0h_{jk}=\dot{\chi}(t)(\epsilon_\times\sigma_{1jk}+\epsilon_+\sigma_{3jk})~.
\end{equation}
In the above $\epsilon_+$ and $\epsilon_\times$ are plus and cross polarization vectors of GWs. $\sigma$,s are Pauli matrices.
Then using \eqref{A.2} in \eqref{H1} one finds
\begin{equation}
    H(t)=\sum_{j=1,2}\left(\frac{p_j^2}{2m}+\sum_{k=1,2}\dot{\chi}(t)\epsilon_a\sigma^a_{jk}x^kp_j+\frac{1}{2}m\omega^2x^2_j\right)
\end{equation}
where $\epsilon_a=(\epsilon_\times,0,\epsilon_+)$.
Expanding the above now one finds
\begin{equation}
\label{A.4}
 H(t)=\sum_{j=1,2}\left(\alpha p_j^2+\beta x^2_j\right)+\gamma(x_1p_1+p_1x_1)-\gamma(x_2p_2+p_2x_2)+\delta(x_1p_2+x_2p_1)~,
\end{equation}
where $\alpha=\frac{1}{2m}$, $\beta=\frac{1}{2}m\omega^2$, $\gamma=\frac{\dot{\chi}(t)}{2}\epsilon_+$, $\delta=\dot{\chi}(t)\epsilon_\times$.
Finally, applying a phase space transformation of the form $x'_i=U_{ij}x_j$ and $p'_i=U_{ij}p_j$ with 
\begin{equation}\label{A.5}
U=
    \begin{pmatrix}
        \cos{\theta}&\sin{\theta}\\
        -\sin{\theta}&\cos{\theta}\\
    \end{pmatrix}~,
\end{equation}
 and then choosing $\tan 2\theta=- \frac{2\gamma}{\delta}$, we find Eq. (\ref{H2}).

\section{Derivation of Eq. (\ref{A2}) and Eq. (\ref{A3})} \label{App2}
Using \eqref{3.3}, \eqref{3.5} in \eqref{3.4} we find
\begin{multline*}
    U_I(t,0) =1+i\Big\{\left(-i\int^t_0 dt_{1} g(t_1)e^{i\Omega t_1}\right)S_++\left(i\int^t_0 dt_{1} g(t_1)e^{-i\Omega t_1}\right)S_-\Big\}\\+i\left(2\int^t_0dt_1\int^{t_1}_0dt_2g(t_1)g(t_2)\sin[{\Omega(t_1-t_2)}]\right)S_0\\-\frac{1}{2}\left(i\int^t_0 dt_{1} g(t_1)e^{i\Omega t_1}S_+-i\int^t_0 dt_{1} g(t_1)e^{-i\Omega t_1}S_-+i\int^t_0 dt_{1} g(t_1)e^{i\Omega t_1}\Tilde{S}_--i\int^t_0 dt_{1} g(t_1)e^{-i\Omega t_1}S_+\right)^2\\+i\Big\{\left(-i\int^t_0 dt_{1} g(t_1)e^{i\Omega t_1}\right)\Tilde{S}_-+\left(i\int^t_0 dt_{1} g(t_1)e^{-i\Omega t_1}\right)\Tilde{S}_+\Big\}\\-i\left(2\int^t_0dt_1\int^{t_1}_0dt_2g(t_1)g(t_2)\sin[{\Omega(t_1-t_2)}]\right)\Tilde{S}_0~.
\end{multline*}
Next using \eqref{j}, one obtains the required Eq. (\ref{A2}).

Now, in order to get \eqref{A3}, we write the effect of $U_I(t,0)$ on the state $\ket{0,0;\beta,t=0}$. It consists of several parts. All terms are calculated separately in the following manner. The first term we calculate as
\begin{multline}\label{24}
  i(c_0S_0-c_+S_++c_-S_-)\ket{0,0;\beta,t=0}= (1-e^{-\beta\omega})\sum_{n_1,n_2=0}^\infty e^{-n_1\beta\omega/2}e^{-n_2\beta\omega/2}\\i(c_0S_0-c_+S_++c_-S_-)\ket{n_1,n_2,\Tilde{n}_1,\Tilde{n}_2}~,
\end{multline}
where
    \begin{equation}\label{25}
\begin{split}
    (c_0S_0-c_+S_++c_-S_-)\ket{n_1,n_2,\Tilde{n}_1,\Tilde{n}_2}=\frac{c_0}{2}(n_1+n_2)\ket{n_1,n_2,\Tilde{n}_1,\Tilde{n}_2}\\-c_+\sqrt{(n_1+1)(n_2+1)}\ket{n_1+1,n_2+1,\Tilde{n}_1,\Tilde{n}_2}\\+c_-\sqrt{n_1n_2}\ket{n_1-1,n_2-1,\Tilde{n}_1,\Tilde{n}_2}~.
    \end{split}
\end{equation}
Similarly the second term is given by
\begin{multline}\label{26}
  i(c_-\Tilde{S}+-c_+\Tilde{S}_--c_0\Tilde{S}_0)\ket{0,0;\beta,t=0}= (1-e^{-\beta\omega})\sum_{n_1,n_2=0}^\infty e^{-n_1\beta\omega/2}e^{-n_2\beta\omega/2}\\i(c_-\Tilde{S}+-c_+\Tilde{S}_--c_0\Tilde{S}_0)\ket{n_1,n_2,\Tilde{n}_1,\Tilde{n}_2}~, 
\end{multline}
where
\begin{equation}\label{27}
    \begin{split}
       (c_-\Tilde{S}+-c_+\Tilde{S}_--c_0\Tilde{S}_0)\ket{n_1,n_2,\Tilde{n}_1,\Tilde{n}_2}=c_-\sqrt{(n_1+1)(n_2+1)}\ket{n_1,n_2,\Tilde{n}_1+1,\Tilde{n}_2+1}\\-c_+\sqrt{n_1n_2}\ket{n_1,n_2,\Tilde{n}_1-1,\Tilde{n}_2-1}\\-\frac{c_0}{2}(n_1+n_2)\ket{n_1,n_2,\Tilde{n}_1,\Tilde{n}_2}~.
    \end{split}
\end{equation}
Lastly we find
\begin{multline}\label{28}
    (c_+S_+-c_-S_-+c_+\Tilde{S}_--c_-\Tilde{S}_+)^2\ket{n_1,n_2,\Tilde{n}_1,\Tilde{n}_2}=\\(c_+)^2\sqrt{(n_1+1)(n_1+2)(n_2+1)(n_2+2)}\ket{n_1+2,n_2+2,\Tilde{n}_1,\Tilde{n}_2}-2c_-c_+(n_1+1)(n_2+1)\ket{n_1,n_2,\Tilde{n}_1,\Tilde{n}_2}\\+2(c_+)^2\sqrt{(n_1+1)n_1(n_2+1)n_2}\ket{n_1+1,n_2+1,\Tilde{n}_1-1,\Tilde{n}_2-1}\\-2c_-c_+(n_1+1)(n_2+1)\ket{n_1+1,n_2+1,\Tilde{n}_1+1,\Tilde{n}_2+1}-2c_-c_+n_1n_2\ket{n_1,n_2,\Tilde{n}_1,\Tilde{n}_2}\\+(c_-)^2\sqrt{(n_1-1)n_1(n_2-1)n_2}\ket{n_1-2,n_2-2,\Tilde{n}_1,\Tilde{n}_2}-2c_-c_+n_1n_2\ket{n_1-1,n_2-1,\Tilde{n}_1-1,\Tilde{n}_2-1}
    \\+2(c_-)^2\sqrt{(n_1+1)n_1(n_2+1)n_2}\ket{n_1-1,n_2-1,\Tilde{n}_1+1,\Tilde{n}_2+1}\\+(c_+)^2\sqrt{(n_1-1)n_1(n_2-1)n_2}\ket{n_1,n_2,\Tilde{n}_1-2,\Tilde{n}_2-2}\\+(c_-)^2\sqrt{(n_1+1)(n_1+2)(n_2+1)(n_2+2)}\ket{n_1,n_2,\Tilde{n}_1+2,\Tilde{n}_2+2}~.
\end{multline}
Finally accumulating all these in $\ket{0,0;\beta,t}_{I} = U_I(t,0) \ket{0,0;\beta,t=0}$, we reach at Eq. (\ref{A3}).
\end{appendix}

\bibliographystyle{apsrev}
\bibliography{bibtexfile1.bib}

\begin{thebibliography}{70}
\expandafter\ifx\csname natexlab\endcsname\relax\def\natexlab#1{#1}\fi
\expandafter\ifx\csname bibnamefont\endcsname\relax
  \def\bibnamefont#1{#1}\fi
\expandafter\ifx\csname bibfnamefont\endcsname\relax
  \def\bibfnamefont#1{#1}\fi
\expandafter\ifx\csname citenamefont\endcsname\relax
  \def\citenamefont#1{#1}\fi
\expandafter\ifx\csname url\endcsname\relax
  \def\url#1{\texttt{#1}}\fi
\expandafter\ifx\csname urlprefix\endcsname\relax\def\urlprefix{URL }\fi
\providecommand{\bibinfo}[2]{#2}
\providecommand{\eprint}[2][]{\url{#2}}

\bibitem[{\citenamefont{Abbott et~al.}(2016)}]{Abbott:2016blz}
\bibinfo{author}{\bibfnamefont{B.~P.} \bibnamefont{Abbott}} \bibnamefont{et~al.} (\bibinfo{collaboration}{Virgo, LIGO Scientific}), \bibinfo{journal}{Phys. Rev. Lett.} \textbf{\bibinfo{volume}{116}}, \bibinfo{pages}{061102} (\bibinfo{year}{2016}), \eprint{arXiv:1602.03837}.

\bibitem[{\citenamefont{Dokuchaev et~al.}(2009)\citenamefont{Dokuchaev, Eroshenko, and Rubin}}]{Dokuchaev:2009}
\bibinfo{author}{\bibfnamefont{V.~I.} \bibnamefont{Dokuchaev}}, \bibinfo{author}{\bibfnamefont{Y.~N.} \bibnamefont{Eroshenko}}, \bibnamefont{and} \bibinfo{author}{\bibfnamefont{S.~G.} \bibnamefont{Rubin}}, \bibinfo{journal}{Astronomy Letters} \textbf{\bibinfo{volume}{35}}, \bibinfo{pages}{143} (\bibinfo{year}{2009}).

\bibitem[{\citenamefont{Lasky}(2015)}]{Lasky2015}
\bibinfo{author}{\bibfnamefont{P.~D.} \bibnamefont{Lasky}}, \bibinfo{journal}{Publications of the Astronomical Society of Australia} \textbf{\bibinfo{volume}{32}}, \bibinfo{pages}{e034} (\bibinfo{year}{2015}), \urlprefix\url{https://doi.org/10.1017/pasa.2015.35}.

\bibitem[{\citenamefont{Boyle and Steinhardt}(2008)}]{Boyle:2008rg}
\bibinfo{author}{\bibfnamefont{L.~A.} \bibnamefont{Boyle}} \bibnamefont{and} \bibinfo{author}{\bibfnamefont{P.~J.} \bibnamefont{Steinhardt}}, \bibinfo{journal}{Phys. Rev. D} \textbf{\bibinfo{volume}{77}}, \bibinfo{pages}{063504} (\bibinfo{year}{2008}), \eprint{astro-ph/0512014}.

\bibitem[{\citenamefont{Kiefer}(2004)}]{Kiefer2004}
\bibinfo{author}{\bibfnamefont{C.}~\bibnamefont{Kiefer}}, \emph{\bibinfo{title}{Quantum Gravity}}, vol. \bibinfo{volume}{124} of \emph{\bibinfo{series}{International Series of Monographs on Physics}} (\bibinfo{publisher}{Clarendon Press}, \bibinfo{address}{Oxford}, \bibinfo{year}{2004}), ISBN \bibinfo{isbn}{978-0-19-958520-5}.

\bibitem[{\citenamefont{Aasi and et~al.}(2015)}]{Aasi2015}
\bibinfo{author}{\bibfnamefont{J.}~\bibnamefont{Aasi}} \bibnamefont{and} \bibinfo{author}{\bibnamefont{et~al.}}, \bibinfo{journal}{Classical and Quantum Gravity} \textbf{\bibinfo{volume}{32}}, \bibinfo{pages}{074001} (\bibinfo{year}{2015}).

\bibitem[{\citenamefont{Acernese and et~al.}(2014)}]{Acernese2014}
\bibinfo{author}{\bibfnamefont{F.}~\bibnamefont{Acernese}} \bibnamefont{and} \bibinfo{author}{\bibnamefont{et~al.}}, \bibinfo{journal}{Classical and Quantum Gravity} \textbf{\bibinfo{volume}{32}}, \bibinfo{pages}{024001} (\bibinfo{year}{2014}).

\bibitem[{\citenamefont{Aso et~al.}(2013)\citenamefont{Aso, Michimura, Somiya, Ando, Miyakawa, Sekiguchi, Tatsumi, and Yamamoto}}]{Aso2013}
\bibinfo{author}{\bibfnamefont{Y.}~\bibnamefont{Aso}}, \bibinfo{author}{\bibfnamefont{Y.}~\bibnamefont{Michimura}}, \bibinfo{author}{\bibfnamefont{K.}~\bibnamefont{Somiya}}, \bibinfo{author}{\bibfnamefont{M.}~\bibnamefont{Ando}}, \bibinfo{author}{\bibfnamefont{O.}~\bibnamefont{Miyakawa}}, \bibinfo{author}{\bibfnamefont{T.}~\bibnamefont{Sekiguchi}}, \bibinfo{author}{\bibfnamefont{D.}~\bibnamefont{Tatsumi}}, \bibnamefont{and} \bibinfo{author}{\bibfnamefont{H.}~\bibnamefont{Yamamoto}}, \bibinfo{journal}{Physical Review D} \textbf{\bibinfo{volume}{88}}, \bibinfo{pages}{043007} (\bibinfo{year}{2013}).

\bibitem[{\citenamefont{Aso and et~al. (The KAGRA~Collaboration)}(2013)}]{AsoKAGRA2013}
\bibinfo{author}{\bibfnamefont{Y.}~\bibnamefont{Aso}} \bibnamefont{and} \bibinfo{author}{\bibnamefont{et~al. (The KAGRA~Collaboration)}}, \bibinfo{journal}{Physical Review D} \textbf{\bibinfo{volume}{88}}, \bibinfo{pages}{043007} (\bibinfo{year}{2013}).

\bibitem[{\citenamefont{Abbott et~al.}(2023)\citenamefont{Abbott, et~al. (LIGO Scientific~Collaboration, and Collaboration)}}]{Abbott2023}
\bibinfo{author}{\bibfnamefont{B.}~\bibnamefont{Abbott}}, \bibinfo{author}{\bibfnamefont{V.~C.} \bibnamefont{et~al. (LIGO Scientific~Collaboration}}, \bibnamefont{and} \bibinfo{author}{\bibfnamefont{K.}~\bibnamefont{Collaboration)}}, \bibinfo{journal}{Physical Review X} \textbf{\bibinfo{volume}{13}}, \bibinfo{pages}{041039} (\bibinfo{year}{2023}).

\bibitem[{\citenamefont{Seto et~al.}(2001)\citenamefont{Seto, Kawamura, and Nakamura}}]{Seto2001}
\bibinfo{author}{\bibfnamefont{N.}~\bibnamefont{Seto}}, \bibinfo{author}{\bibfnamefont{S.}~\bibnamefont{Kawamura}}, \bibnamefont{and} \bibinfo{author}{\bibfnamefont{T.}~\bibnamefont{Nakamura}}, \bibinfo{journal}{Physical Review Letters} \textbf{\bibinfo{volume}{87}}, \bibinfo{pages}{221103} (\bibinfo{year}{2001}).

\bibitem[{\citenamefont{et~al.}(2006)}]{Kawamura2006}
\bibinfo{author}{\bibfnamefont{S.~K.} \bibnamefont{et~al.}}, \bibinfo{journal}{Classical and Quantum Gravity} \textbf{\bibinfo{volume}{23}}, \bibinfo{pages}{S125} (\bibinfo{year}{2006}).

\bibitem[{\citenamefont{Yagi and Seto}(2011)}]{Yagi2011}
\bibinfo{author}{\bibfnamefont{K.}~\bibnamefont{Yagi}} \bibnamefont{and} \bibinfo{author}{\bibfnamefont{N.}~\bibnamefont{Seto}}, \bibinfo{journal}{Physical Review D} \textbf{\bibinfo{volume}{83}}, \bibinfo{pages}{044011} (\bibinfo{year}{2011}).

\bibitem[{\citenamefont{Yagi and Seto}(2017)}]{Yagi2017}
\bibinfo{author}{\bibfnamefont{K.}~\bibnamefont{Yagi}} \bibnamefont{and} \bibinfo{author}{\bibfnamefont{N.}~\bibnamefont{Seto}}, \bibinfo{journal}{Physical Review D} \textbf{\bibinfo{volume}{95}}, \bibinfo{pages}{109901(E)} (\bibinfo{year}{2017}).

\bibitem[{Qua(2021)}]{QuantumSciTech2021}
\bibinfo{journal}{Quantum Science and Technology} \textbf{\bibinfo{volume}{6}}, \bibinfo{pages}{044003} (\bibinfo{year}{2021}).

\bibitem[{\citenamefont{Kramer and Champion}(2013)}]{Kramer2013}
\bibinfo{author}{\bibfnamefont{M.}~\bibnamefont{Kramer}} \bibnamefont{and} \bibinfo{author}{\bibfnamefont{D.~J.} \bibnamefont{Champion}}, \bibinfo{journal}{Classical and Quantum Gravity} \textbf{\bibinfo{volume}{30}}, \bibinfo{pages}{224009} (\bibinfo{year}{2013}).

\bibitem[{\citenamefont{et~al.}(2016)}]{Babak2016}
\bibinfo{author}{\bibfnamefont{S.~B.} \bibnamefont{et~al.}}, \bibinfo{journal}{Monthly Notices of the Royal Astronomical Society} \textbf{\bibinfo{volume}{455}}, \bibinfo{pages}{1665} (\bibinfo{year}{2016}).

\bibitem[{\citenamefont{et~al.}(2015{\natexlab{a}})}]{Lentati2015}
\bibinfo{author}{\bibfnamefont{L.~L.} \bibnamefont{et~al.}}, \bibinfo{journal}{Monthly Notices of the Royal Astronomical Society} \textbf{\bibinfo{volume}{453}}, \bibinfo{pages}{2576} (\bibinfo{year}{2015}{\natexlab{a}}).

\bibitem[{\citenamefont{et~al.}(2015{\natexlab{b}})}]{Shannon2015}
\bibinfo{author}{\bibfnamefont{R.~M.~S.} \bibnamefont{et~al.}}, \bibinfo{journal}{Science} \textbf{\bibinfo{volume}{349}}, \bibinfo{pages}{1522} (\bibinfo{year}{2015}{\natexlab{b}}).

\bibitem[{\citenamefont{et~al.}(2013)}]{Manchester2013}
\bibinfo{author}{\bibfnamefont{R.~N.~M.} \bibnamefont{et~al.}}, \bibinfo{journal}{Publications of the Astronomical Society of Australia} \textbf{\bibinfo{volume}{30}}, \bibinfo{pages}{17} (\bibinfo{year}{2013}).

\bibitem[{\citenamefont{Hobbs et~al.}(2010)}]{Hobbs2010}
\bibinfo{author}{\bibfnamefont{G.}~\bibnamefont{Hobbs}} \bibnamefont{et~al.}, \bibinfo{journal}{Classical and Quantum Gravity} \textbf{\bibinfo{volume}{27}}, \bibinfo{pages}{084013} (\bibinfo{year}{2010}).

\bibitem[{\citenamefont{Verbiest et~al.}(2016)}]{Verbiest2016}
\bibinfo{author}{\bibfnamefont{J.~P.~W.} \bibnamefont{Verbiest}} \bibnamefont{et~al.}, \bibinfo{journal}{Monthly Notices of the Royal Astronomical Society} \textbf{\bibinfo{volume}{458}}, \bibinfo{pages}{1267} (\bibinfo{year}{2016}).

\bibitem[{\citenamefont{Hazboun et~al.}(2018)\citenamefont{Hazboun, Mingarelli, and Lee}}]{Hazboun2018}
\bibinfo{author}{\bibfnamefont{J.~S.} \bibnamefont{Hazboun}}, \bibinfo{author}{\bibfnamefont{C.~M.~F.} \bibnamefont{Mingarelli}}, \bibnamefont{and} \bibinfo{author}{\bibfnamefont{K.}~\bibnamefont{Lee}} (\bibinfo{year}{2018}), \eprint{1810.10527}.

\bibitem[{\citenamefont{Agullo et~al.}(2021)\citenamefont{Agullo, Cardoso, del Rio, Maggiore, and Pullin}}]{PhysRevLett.126.041302}
\bibinfo{author}{\bibfnamefont{I.}~\bibnamefont{Agullo}}, \bibinfo{author}{\bibfnamefont{V.}~\bibnamefont{Cardoso}}, \bibinfo{author}{\bibfnamefont{A.}~\bibnamefont{del Rio}}, \bibinfo{author}{\bibfnamefont{M.}~\bibnamefont{Maggiore}}, \bibnamefont{and} \bibinfo{author}{\bibfnamefont{J.}~\bibnamefont{Pullin}}, \bibinfo{journal}{Phys. Rev. Lett.} \textbf{\bibinfo{volume}{126}}, \bibinfo{pages}{041302} (\bibinfo{year}{2021}), \urlprefix\url{https://link.aps.org/doi/10.1103/PhysRevLett.126.041302}.

\bibitem[{\citenamefont{Page and Geilker}(1981)}]{PhysRevLett.47.979}
\bibinfo{author}{\bibfnamefont{D.~N.} \bibnamefont{Page}} \bibnamefont{and} \bibinfo{author}{\bibfnamefont{C.~D.} \bibnamefont{Geilker}}, \bibinfo{journal}{Phys. Rev. Lett.} \textbf{\bibinfo{volume}{47}}, \bibinfo{pages}{979} (\bibinfo{year}{1981}), \urlprefix\url{https://link.aps.org/doi/10.1103/PhysRevLett.47.979}.

\bibitem[{\citenamefont{Parikh et~al.}(2021{\natexlab{a}})\citenamefont{Parikh, Wilczek, and Zahariade}}]{PhysRevLett.127.081602}
\bibinfo{author}{\bibfnamefont{M.}~\bibnamefont{Parikh}}, \bibinfo{author}{\bibfnamefont{F.}~\bibnamefont{Wilczek}}, \bibnamefont{and} \bibinfo{author}{\bibfnamefont{G.}~\bibnamefont{Zahariade}}, \bibinfo{journal}{Phys. Rev. Lett.} \textbf{\bibinfo{volume}{127}}, \bibinfo{pages}{081602} (\bibinfo{year}{2021}{\natexlab{a}}), \urlprefix\url{https://link.aps.org/doi/10.1103/PhysRevLett.127.081602}.

\bibitem[{\citenamefont{Kahn et~al.}(2024)\citenamefont{Kahn, Sch\"utte-Engel, and Trickle}}]{PhysRevD.109.096023}
\bibinfo{author}{\bibfnamefont{Y.}~\bibnamefont{Kahn}}, \bibinfo{author}{\bibfnamefont{J.}~\bibnamefont{Sch\"utte-Engel}}, \bibnamefont{and} \bibinfo{author}{\bibfnamefont{T.}~\bibnamefont{Trickle}}, \bibinfo{journal}{Phys. Rev. D} \textbf{\bibinfo{volume}{109}}, \bibinfo{pages}{096023} (\bibinfo{year}{2024}), \urlprefix\url{https://link.aps.org/doi/10.1103/PhysRevD.109.096023}.

\bibitem[{\citenamefont{Parikh et~al.}(2021{\natexlab{b}})\citenamefont{Parikh, Wilczek, and Zahariade}}]{PhysRevD.104.046021}
\bibinfo{author}{\bibfnamefont{M.}~\bibnamefont{Parikh}}, \bibinfo{author}{\bibfnamefont{F.}~\bibnamefont{Wilczek}}, \bibnamefont{and} \bibinfo{author}{\bibfnamefont{G.}~\bibnamefont{Zahariade}}, \bibinfo{journal}{Phys. Rev. D} \textbf{\bibinfo{volume}{104}}, \bibinfo{pages}{046021} (\bibinfo{year}{2021}{\natexlab{b}}), \urlprefix\url{https://link.aps.org/doi/10.1103/PhysRevD.104.046021}.

\bibitem[{\citenamefont{Jacobson}(1995)}]{PhysRevLett.75.1260}
\bibinfo{author}{\bibfnamefont{T.}~\bibnamefont{Jacobson}}, \bibinfo{journal}{Phys. Rev. Lett.} \textbf{\bibinfo{volume}{75}}, \bibinfo{pages}{1260} (\bibinfo{year}{1995}), \urlprefix\url{https://link.aps.org/doi/10.1103/PhysRevLett.75.1260}.

\bibitem[{\citenamefont{Tobar et~al.}(2024)\citenamefont{Tobar, Manikandan, Beitel et~al.}}]{Tobar2024}
\bibinfo{author}{\bibfnamefont{G.}~\bibnamefont{Tobar}}, \bibinfo{author}{\bibfnamefont{S.~K.} \bibnamefont{Manikandan}}, \bibinfo{author}{\bibfnamefont{T.}~\bibnamefont{Beitel}}, \bibnamefont{et~al.}, \bibinfo{journal}{Nature Communications} \textbf{\bibinfo{volume}{15}}, \bibinfo{pages}{7229} (\bibinfo{year}{2024}).

\bibitem[{\citenamefont{Clauser}(1974)}]{Clauser1974}
\bibinfo{author}{\bibfnamefont{J.~F.} \bibnamefont{Clauser}}, \bibinfo{journal}{Physical Review D} \textbf{\bibinfo{volume}{9}}, \bibinfo{pages}{853} (\bibinfo{year}{1974}).

\bibitem[{\citenamefont{Carney et~al.}(2024)\citenamefont{Carney, Domcke, and Rodd}}]{PhysRevD.109.044009}
\bibinfo{author}{\bibfnamefont{D.}~\bibnamefont{Carney}}, \bibinfo{author}{\bibfnamefont{V.}~\bibnamefont{Domcke}}, \bibnamefont{and} \bibinfo{author}{\bibfnamefont{N.~L.} \bibnamefont{Rodd}}, \bibinfo{journal}{Phys. Rev. D} \textbf{\bibinfo{volume}{109}}, \bibinfo{pages}{044009} (\bibinfo{year}{2024}), \urlprefix\url{https://link.aps.org/doi/10.1103/PhysRevD.109.044009}.

\bibitem[{\citenamefont{Bose et~al.}(2017)\citenamefont{Bose, Mazumdar, Morley, Ulbricht, Toros, Paternostro, Geraci, Barker, Kim, and Milburn}}]{Bose2017}
\bibinfo{author}{\bibfnamefont{S.}~\bibnamefont{Bose}}, \bibinfo{author}{\bibfnamefont{A.}~\bibnamefont{Mazumdar}}, \bibinfo{author}{\bibfnamefont{G.~W.} \bibnamefont{Morley}}, \bibinfo{author}{\bibfnamefont{H.}~\bibnamefont{Ulbricht}}, \bibinfo{author}{\bibfnamefont{M.}~\bibnamefont{Toros}}, \bibinfo{author}{\bibfnamefont{M.}~\bibnamefont{Paternostro}}, \bibinfo{author}{\bibfnamefont{A.~A.} \bibnamefont{Geraci}}, \bibinfo{author}{\bibfnamefont{P.~F.} \bibnamefont{Barker}}, \bibinfo{author}{\bibfnamefont{M.~S.} \bibnamefont{Kim}}, \bibnamefont{and} \bibinfo{author}{\bibfnamefont{G.}~\bibnamefont{Milburn}}, \bibinfo{journal}{Physical Review Letters} \textbf{\bibinfo{volume}{119}}, \bibinfo{pages}{240401} (\bibinfo{year}{2017}).

\bibitem[{\citenamefont{Marletto and Vedral}(2017)}]{Marletto2017}
\bibinfo{author}{\bibfnamefont{C.}~\bibnamefont{Marletto}} \bibnamefont{and} \bibinfo{author}{\bibfnamefont{V.}~\bibnamefont{Vedral}}, \bibinfo{journal}{Physical Review Letters} \textbf{\bibinfo{volume}{119}}, \bibinfo{pages}{240402} (\bibinfo{year}{2017}).

\bibitem[{\citenamefont{Nandi and Majhi}(2024)}]{Nandi:2024jyf}
\bibinfo{author}{\bibfnamefont{P.}~\bibnamefont{Nandi}} \bibnamefont{and} \bibinfo{author}{\bibfnamefont{B.~R.} \bibnamefont{Majhi}}, \bibinfo{journal}{Phys. Lett. B} \textbf{\bibinfo{volume}{857}}, \bibinfo{pages}{138988} (\bibinfo{year}{2024}), \eprint{2403.11253}.

\bibitem[{\citenamefont{Nandi et~al.}(2024)\citenamefont{Nandi, Majhi, Debnath, and Kala}}]{Nandi:2024zxp}
\bibinfo{author}{\bibfnamefont{P.}~\bibnamefont{Nandi}}, \bibinfo{author}{\bibfnamefont{B.~R.} \bibnamefont{Majhi}}, \bibinfo{author}{\bibfnamefont{N.}~\bibnamefont{Debnath}}, \bibnamefont{and} \bibinfo{author}{\bibfnamefont{S.}~\bibnamefont{Kala}}, \bibinfo{journal}{Phys. Lett. B} \textbf{\bibinfo{volume}{853}}, \bibinfo{pages}{138706} (\bibinfo{year}{2024}), \eprint{2401.02778}.

\bibitem[{\citenamefont{Ruggiero}(2025)}]{Ruggiero:2024pzv}
\bibinfo{author}{\bibfnamefont{M.~L.} \bibnamefont{Ruggiero}}, \bibinfo{journal}{Gen. Rel. Grav.} \textbf{\bibinfo{volume}{57}}, \bibinfo{pages}{52} (\bibinfo{year}{2025}), \eprint{2407.15968}.

\bibitem[{\citenamefont{Kaku and Nambu}(2025)}]{Kaku:2024lgs}
\bibinfo{author}{\bibfnamefont{Y.}~\bibnamefont{Kaku}} \bibnamefont{and} \bibinfo{author}{\bibfnamefont{Y.}~\bibnamefont{Nambu}}, \bibinfo{journal}{Phys. Rev. D} \textbf{\bibinfo{volume}{111}}, \bibinfo{pages}{046026} (\bibinfo{year}{2025}), \eprint{2411.12997}.

\bibitem[{\citenamefont{Jones et~al.}(2024)\citenamefont{Jones, Bailey, Gretarsson, and Poon}}]{Jones:2024npd}
\bibinfo{author}{\bibfnamefont{P.}~\bibnamefont{Jones}}, \bibinfo{author}{\bibfnamefont{Q.~G.} \bibnamefont{Bailey}}, \bibinfo{author}{\bibfnamefont{A.}~\bibnamefont{Gretarsson}}, \bibnamefont{and} \bibinfo{author}{\bibfnamefont{E.}~\bibnamefont{Poon}} (\bibinfo{year}{2024}), \eprint{2411.15632}.

\bibitem[{\citenamefont{Mart\'in-Mart\'inez and Perche}(2023)}]{Martin-Martinez:2023}
\bibinfo{author}{\bibfnamefont{E.}~\bibnamefont{Mart\'in-Mart\'inez}} \bibnamefont{and} \bibinfo{author}{\bibfnamefont{T.~R.} \bibnamefont{Perche}}, \bibinfo{journal}{Phys. Rev. D} \textbf{\bibinfo{volume}{108}} (\bibinfo{year}{2023}), \urlprefix\url{https://link.aps.org/doi/10.1103/PhysRevD.108.L101702}.

\bibitem[{\citenamefont{Speliotopoulos}(1995)}]{Speliotopoulos1995}
\bibinfo{author}{\bibfnamefont{A.~D.} \bibnamefont{Speliotopoulos}}, \bibinfo{journal}{Physical Review D} \textbf{\bibinfo{volume}{51}}, \bibinfo{pages}{1701} (\bibinfo{year}{1995}), \urlprefix\url{https://link.aps.org/doi/10.1103/PhysRevD.51.1701}.

\bibitem[{\citenamefont{Nandi et~al.}(2023)\citenamefont{Nandi, Pal, Pal, and Majhi}}]{PhysRevD.108.124069}
\bibinfo{author}{\bibfnamefont{P.}~\bibnamefont{Nandi}}, \bibinfo{author}{\bibfnamefont{S.}~\bibnamefont{Pal}}, \bibinfo{author}{\bibfnamefont{S.~K.} \bibnamefont{Pal}}, \bibnamefont{and} \bibinfo{author}{\bibfnamefont{B.~R.} \bibnamefont{Majhi}}, \bibinfo{journal}{Phys. Rev. D} \textbf{\bibinfo{volume}{108}}, \bibinfo{pages}{124069} (\bibinfo{year}{2023}), \urlprefix\url{https://link.aps.org/doi/10.1103/PhysRevD.108.124069}.

\bibitem[{\citenamefont{Colella et~al.}(1975)\citenamefont{Colella, Overhauser, and Werner}}]{Colella1975}
\bibinfo{author}{\bibfnamefont{R.}~\bibnamefont{Colella}}, \bibinfo{author}{\bibfnamefont{A.~W.} \bibnamefont{Overhauser}}, \bibnamefont{and} \bibinfo{author}{\bibfnamefont{S.~A.} \bibnamefont{Werner}}, \bibinfo{journal}{Physical Review Letters} \textbf{\bibinfo{volume}{34}}, \bibinfo{pages}{1472} (\bibinfo{year}{1975}), \urlprefix\url{https://link.aps.org/doi/10.1103/PhysRevLett.34.1472}.

\bibitem[{\citenamefont{Pikovski et~al.}(2015)\citenamefont{Pikovski, Zych, Costa, and Brukner}}]{Pikovski2015}
\bibinfo{author}{\bibfnamefont{I.}~\bibnamefont{Pikovski}}, \bibinfo{author}{\bibfnamefont{M.}~\bibnamefont{Zych}}, \bibinfo{author}{\bibfnamefont{F.}~\bibnamefont{Costa}}, \bibnamefont{and} \bibinfo{author}{\bibfnamefont{C.}~\bibnamefont{Brukner}}, \bibinfo{journal}{Nature Physics} \textbf{\bibinfo{volume}{11}}, \bibinfo{pages}{668} (\bibinfo{year}{2015}), \eprint{1311.1095}.

\bibitem[{\citenamefont{Caves et~al.}(1980)\citenamefont{Caves, Thorne, Drever, Sandberg, and Zimmermann}}]{Caves1980}
\bibinfo{author}{\bibfnamefont{C.~M.} \bibnamefont{Caves}}, \bibinfo{author}{\bibfnamefont{K.~S.} \bibnamefont{Thorne}}, \bibinfo{author}{\bibfnamefont{R.~W.~P.} \bibnamefont{Drever}}, \bibinfo{author}{\bibfnamefont{V.~D.} \bibnamefont{Sandberg}}, \bibnamefont{and} \bibinfo{author}{\bibfnamefont{M.}~\bibnamefont{Zimmermann}}, \bibinfo{journal}{Reviews of Modern Physics} \textbf{\bibinfo{volume}{52}}, \bibinfo{pages}{341} (\bibinfo{year}{1980}), \urlprefix\url{https://link.aps.org/doi/10.1103/RevModPhys.52.341}.

\bibitem[{\citenamefont{Thorne}(1997)}]{Thorne1997}
\bibinfo{author}{\bibfnamefont{K.~S.} \bibnamefont{Thorne}} (\bibinfo{year}{1997}), \eprint{gr-qc/9704042}.

\bibitem[{\citenamefont{Dimopoulos et~al.}(2008)\citenamefont{Dimopoulos, Graham, Hogan, Kasevich, and Rajendran}}]{Dimopoulos2008}
\bibinfo{author}{\bibfnamefont{S.}~\bibnamefont{Dimopoulos}}, \bibinfo{author}{\bibfnamefont{P.~W.} \bibnamefont{Graham}}, \bibinfo{author}{\bibfnamefont{J.~M.} \bibnamefont{Hogan}}, \bibinfo{author}{\bibfnamefont{M.~A.} \bibnamefont{Kasevich}}, \bibnamefont{and} \bibinfo{author}{\bibfnamefont{S.}~\bibnamefont{Rajendran}}, \bibinfo{journal}{Physical Review D} \textbf{\bibinfo{volume}{78}}, \bibinfo{pages}{122002} (\bibinfo{year}{2008}), \urlprefix\url{https://link.aps.org/doi/10.1103/PhysRevD.78.122002}.

\bibitem[{\citenamefont{Das}(1997)}]{Das:1997gg}
\bibinfo{author}{\bibfnamefont{A.~K.} \bibnamefont{Das}}, \emph{\bibinfo{title}{{Finite Temperature Field Theory}}} (\bibinfo{publisher}{World Scientific}, \bibinfo{address}{New York}, \bibinfo{year}{1997}), ISBN \bibinfo{isbn}{978-981-02-2856-9, 978-981-4498-23-4}.

\bibitem[{\citenamefont{Takahashi and Umezawa}(1996)}]{Takahashi:1996zn}
\bibinfo{author}{\bibfnamefont{Y.}~\bibnamefont{Takahashi}} \bibnamefont{and} \bibinfo{author}{\bibfnamefont{H.}~\bibnamefont{Umezawa}}, \bibinfo{journal}{Int. J. Mod. Phys. B} \textbf{\bibinfo{volume}{10}}, \bibinfo{pages}{1755} (\bibinfo{year}{1996}).

\bibitem[{\citenamefont{Else et~al.}(2016)\citenamefont{Else, Bauer, and Nayak}}]{PhysRevLett.117.090402}
\bibinfo{author}{\bibfnamefont{D.~V.} \bibnamefont{Else}}, \bibinfo{author}{\bibfnamefont{B.}~\bibnamefont{Bauer}}, \bibnamefont{and} \bibinfo{author}{\bibfnamefont{C.}~\bibnamefont{Nayak}}, \bibinfo{journal}{Phys. Rev. Lett.} \textbf{\bibinfo{volume}{117}}, \bibinfo{pages}{090402} (\bibinfo{year}{2016}), \urlprefix\url{https://link.aps.org/doi/10.1103/PhysRevLett.117.090402}.

\bibitem[{\citenamefont{Misner et~al.}(1973)\citenamefont{Misner, Thorne, and Wheeler}}]{Misner:1973prb}
\bibinfo{author}{\bibfnamefont{C.~W.} \bibnamefont{Misner}}, \bibinfo{author}{\bibfnamefont{K.~S.} \bibnamefont{Thorne}}, \bibnamefont{and} \bibinfo{author}{\bibfnamefont{J.~A.} \bibnamefont{Wheeler}}, \emph{\bibinfo{title}{{Gravitation}}} (\bibinfo{publisher}{W. H. Freeman}, \bibinfo{address}{San Francisco}, \bibinfo{year}{1973}), ISBN \bibinfo{isbn}{978-0-7167-0344-0, 978-0-691-17779-3}.

\bibitem[{\citenamefont{Maggiore}(2007)}]{Maggiore:2007ulw}
\bibinfo{author}{\bibfnamefont{M.}~\bibnamefont{Maggiore}}, \emph{\bibinfo{title}{{Gravitational Waves. Vol. 1: Theory and Experiments}}} (\bibinfo{publisher}{Oxford University Press}, \bibinfo{year}{2007}), ISBN \bibinfo{isbn}{978-0-19-171766-6, 978-0-19-852074-0}.

\bibitem[{\citenamefont{Kanno et~al.}(2021)\citenamefont{Kanno, Soda, and Tokuda}}]{PhysRevD.103.044017}
\bibinfo{author}{\bibfnamefont{S.}~\bibnamefont{Kanno}}, \bibinfo{author}{\bibfnamefont{J.}~\bibnamefont{Soda}}, \bibnamefont{and} \bibinfo{author}{\bibfnamefont{J.}~\bibnamefont{Tokuda}}, \bibinfo{journal}{Phys. Rev. D} \textbf{\bibinfo{volume}{103}}, \bibinfo{pages}{044017} (\bibinfo{year}{2021}), \urlprefix\url{https://link.aps.org/doi/10.1103/PhysRevD.103.044017}.

\bibitem[{\citenamefont{Biswas and Das}(1988)}]{Biswas1989}
\bibinfo{author}{\bibfnamefont{S.~N.} \bibnamefont{Biswas}} \bibnamefont{and} \bibinfo{author}{\bibfnamefont{A.}~\bibnamefont{Das}}, \bibinfo{journal}{Modern Physics Letters A} \textbf{\bibinfo{volume}{3}}, \bibinfo{pages}{549} (\bibinfo{year}{1988}).

\bibitem[{\citenamefont{R{\'e}nyi}(1970)}]{Renyi1970}
\bibinfo{author}{\bibfnamefont{A.}~\bibnamefont{R{\'e}nyi}}, \emph{\bibinfo{title}{Probability Theory}} (\bibinfo{publisher}{Academiai Kiado}, \bibinfo{address}{Budapest, Hungary}, \bibinfo{year}{1970}).

\bibitem[{\citenamefont{Pal et~al.}(2021)\citenamefont{Pal, Nandi, Ghosh, Scholtz, and Chakraborty}}]{Pal:2020jvq}
\bibinfo{author}{\bibfnamefont{S.~K.} \bibnamefont{Pal}}, \bibinfo{author}{\bibfnamefont{P.}~\bibnamefont{Nandi}}, \bibinfo{author}{\bibfnamefont{S.}~\bibnamefont{Ghosh}}, \bibinfo{author}{\bibfnamefont{F.~G.} \bibnamefont{Scholtz}}, \bibnamefont{and} \bibinfo{author}{\bibfnamefont{B.}~\bibnamefont{Chakraborty}}, \bibinfo{journal}{Phys. Lett. A} \textbf{\bibinfo{volume}{403}}, \bibinfo{pages}{127397} (\bibinfo{year}{2021}), \eprint{2012.07166}.

\bibitem[{\citenamefont{Casas-Vázquez and Jou}(2003)}]{J2003}
\bibinfo{author}{\bibfnamefont{J.}~\bibnamefont{Casas-Vázquez}} \bibnamefont{and} \bibinfo{author}{\bibfnamefont{D.}~\bibnamefont{Jou}}, \bibinfo{journal}{Reports on Progress in Physics} \textbf{\bibinfo{volume}{66}}, \bibinfo{pages}{1937} (\bibinfo{year}{2003}), \urlprefix\url{https://dx.doi.org/10.1088/0034-4885/66/11/R03}.

\bibitem[{\citenamefont{Kolekar}(2014)}]{Kolekar:2013xua}
\bibinfo{author}{\bibfnamefont{S.}~\bibnamefont{Kolekar}}, \bibinfo{journal}{Phys. Rev. D} \textbf{\bibinfo{volume}{89}}, \bibinfo{pages}{044036} (\bibinfo{year}{2014}), \eprint{1309.3261}.

\bibitem[{\citenamefont{Kolekar and Padmanabhan}(2014)}]{Kolekar:2013hra}
\bibinfo{author}{\bibfnamefont{S.}~\bibnamefont{Kolekar}} \bibnamefont{and} \bibinfo{author}{\bibfnamefont{T.}~\bibnamefont{Padmanabhan}}, \bibinfo{journal}{Phys. Rev. D} \textbf{\bibinfo{volume}{89}}, \bibinfo{pages}{064055} (\bibinfo{year}{2014}), \eprint{1309.4424}.

\bibitem[{\citenamefont{Chowdhury and Majhi}(2022)}]{Chowdhury:2021ieg}
\bibinfo{author}{\bibfnamefont{P.}~\bibnamefont{Chowdhury}} \bibnamefont{and} \bibinfo{author}{\bibfnamefont{B.~R.} \bibnamefont{Majhi}}, \bibinfo{journal}{JHEP} \textbf{\bibinfo{volume}{05}}, \bibinfo{pages}{025} (\bibinfo{year}{2022}), \eprint{2110.11260}.

\bibitem[{\citenamefont{Ghosh and Majhi}(2024)}]{Ghosh:2024mqy}
\bibinfo{author}{\bibfnamefont{D.}~\bibnamefont{Ghosh}} \bibnamefont{and} \bibinfo{author}{\bibfnamefont{B.~R.} \bibnamefont{Majhi}}, \bibinfo{journal}{Phys. Lett. B} \textbf{\bibinfo{volume}{858}}, \bibinfo{pages}{139015} (\bibinfo{year}{2024}), \eprint{2406.13416}.

\bibitem[{\citenamefont{Chowdhury et~al.}(2019)\citenamefont{Chowdhury, Das, Dalui, and Majhi}}]{Chowdhury:2019set}
\bibinfo{author}{\bibfnamefont{C.}~\bibnamefont{Chowdhury}}, \bibinfo{author}{\bibfnamefont{S.}~\bibnamefont{Das}}, \bibinfo{author}{\bibfnamefont{S.}~\bibnamefont{Dalui}}, \bibnamefont{and} \bibinfo{author}{\bibfnamefont{B.~R.} \bibnamefont{Majhi}}, \bibinfo{journal}{Phys. Rev. D} \textbf{\bibinfo{volume}{99}}, \bibinfo{pages}{045021} (\bibinfo{year}{2019}), \eprint{1902.06900}.

\bibitem[{\citenamefont{Barman and Majhi}(2021)}]{Barman:2021oum}
\bibinfo{author}{\bibfnamefont{S.}~\bibnamefont{Barman}} \bibnamefont{and} \bibinfo{author}{\bibfnamefont{B.~R.} \bibnamefont{Majhi}}, \bibinfo{journal}{JHEP} \textbf{\bibinfo{volume}{03}}, \bibinfo{pages}{245} (\bibinfo{year}{2021}), \eprint{2101.08186}.

\bibitem[{\citenamefont{Barman and Majhi}(2024)}]{Barman:2023bpz}
\bibinfo{author}{\bibfnamefont{D.}~\bibnamefont{Barman}} \bibnamefont{and} \bibinfo{author}{\bibfnamefont{B.~R.} \bibnamefont{Majhi}}, \bibinfo{journal}{Annals Phys.} \textbf{\bibinfo{volume}{465}}, \bibinfo{pages}{169678} (\bibinfo{year}{2024}), \eprint{2303.16022}.

\bibitem[{\citenamefont{Kaniadakis}(1994)}]{PhysRevE.49.5111}
\bibinfo{author}{\bibfnamefont{G.}~\bibnamefont{Kaniadakis}}, \bibinfo{journal}{Phys. Rev. E} \textbf{\bibinfo{volume}{49}}, \bibinfo{pages}{5111} (\bibinfo{year}{1994}), \urlprefix\url{https://link.aps.org/doi/10.1103/PhysRevE.49.5111}.

\bibitem[{\citenamefont{Else et~al.}(2017)\citenamefont{Else, Bauer, and Nayak}}]{PhysRevX.7.011026}
\bibinfo{author}{\bibfnamefont{D.~V.} \bibnamefont{Else}}, \bibinfo{author}{\bibfnamefont{B.}~\bibnamefont{Bauer}}, \bibnamefont{and} \bibinfo{author}{\bibfnamefont{C.}~\bibnamefont{Nayak}}, \bibinfo{journal}{Phys. Rev. X} \textbf{\bibinfo{volume}{7}}, \bibinfo{pages}{011026} (\bibinfo{year}{2017}).

\bibitem[{\citenamefont{Wilczek}(2012)}]{PhysRevLett.109.160401}
\bibinfo{author}{\bibfnamefont{F.}~\bibnamefont{Wilczek}}, \bibinfo{journal}{Phys. Rev. Lett.} \textbf{\bibinfo{volume}{109}}, \bibinfo{pages}{160401} (\bibinfo{year}{2012}), \urlprefix\url{https://link.aps.org/doi/10.1103/PhysRevLett.109.160401}.

\bibitem[{\citenamefont{Sacha and Zakrzewski}(2018)}]{Sacha:2017fqe}
\bibinfo{author}{\bibfnamefont{K.}~\bibnamefont{Sacha}} \bibnamefont{and} \bibinfo{author}{\bibfnamefont{J.}~\bibnamefont{Zakrzewski}}, \bibinfo{journal}{Rept. Prog. Phys.} \textbf{\bibinfo{volume}{81}}, \bibinfo{pages}{016401} (\bibinfo{year}{2018}), \eprint{1704.03735}.

\bibitem[{\citenamefont{Bernardini and Bertolami}(2022)}]{bernardini2022emergent}
\bibinfo{author}{\bibfnamefont{A.~E.} \bibnamefont{Bernardini}} \bibnamefont{and} \bibinfo{author}{\bibfnamefont{O.}~\bibnamefont{Bertolami}}, \bibinfo{journal}{Physics Letters B} \textbf{\bibinfo{volume}{835}}, \bibinfo{pages}{137549} (\bibinfo{year}{2022}).

\bibitem[{\citenamefont{Bertolami and Bernardini}(2024)}]{Bertolami:2024tom}
\bibinfo{author}{\bibfnamefont{O.}~\bibnamefont{Bertolami}} \bibnamefont{and} \bibinfo{author}{\bibfnamefont{A.~E.} \bibnamefont{Bernardini}} (\bibinfo{year}{2024}), \eprint{2402.18238}.

\end{thebibliography}

\end{document}